\documentclass[letterpaper,12pt]{article}
\usepackage{osajnl2} 
\usepackage[draft]{hyperref} 
\usepackage{graphicx,amssymb,amsmath,amsthm,cite,cases}
\usepackage{algorithmic}
\usepackage[ruled]{algorithm}  

\def\Del{\Delta}
\usepackage{verbatim}
\usepackage{color,multirow}

\newcommand{\la}{\langle}
\newcommand{\ra}{\rangle}

\def\ll{{\ell}}
\newcommand{\Spann}{{\mbox{\rm{span}}}}

\newtheorem{remark}{Remark}

\def\be{\begin{equation}}
\def\ee{\end{equation}}
\def\ben{\begin{eqnarray}}
\def\een{\end{eqnarray}}

\def\D{\mathcal{D}}
\def\DX{\mathcal{D}^x}
\def\DY{\mathcal{D}^y}
\def\R{\mathbb{R}}
\def\N{\mathbb{N}}

\def\V{\mathbb{V}}
\def\S{\mathbb{S}}
\def\F{\mathrm{F}}

\def\Nx{N_x}
\def\Ny{N_y}

\def\Nxy{{\Nx \times \Ny}}

\def\Mx{M_x}
\def\My{M_y}

\def\dx{{d}^x}
\def\dy{{d}^y}
\def\llx{{\ll^{x}}}
\def\lly{{\ll^{y}}}
\def\mur{\overline{{\mathrm{SR}}}}
\def\tr{\overline{{\mathrm{t}}}}
\def\omur{{\overline{\mur}}}
\def\sigr{\sigma_\mathrm{SR}}
\def\osigr{\overline{\sigma}_{\rm{SR}}}

\def\IA{\vI_{\rm{Approx}}}
\def\I{\rm{I}}

\def\EA{{\rm{Error}_1}}
\def\EP{{\rm{Error}_2}}

\def\vI{\mathbf{I}}
\def\vv{\mathbf{v}}
\def\vR{\mathbf{R}}
\def\vd{\mathbf{d}}
\def\vD{\mathbf{D}}
\def\vc{\mathbf{c}}
\def\vf{\mathbf{f}}
\def\vC{\mathbf{C}}
\def\vA{\mathbf{A}}
\def\vB{\mathbf{B}}
\def\vW{\mathbf{C}}
\def\vRe{\mathbf{R}}

\def\I{\mathrm{I}}
\def\Re{\mathrm{R}}

\def\op{\hat{P}}

\begin{document}
\title{Sparse Representation of Astronomical Images}
\author{Laura~Rebollo-Neira and James Bowley}
\address{Mathematics Department,\\
Aston University,\\
Birmingham B4 7ET, UK}
\topmargin=0mm

\maketitle
\begin{abstract}
Sparse representation of astronomical images is discussed. 
It is shown that a significant gain in sparsity is achieved
when particular mixed dictionaries are used for approximating 
these types of images with greedy selection strategies. 
Experiments are conducted to confirm:
i)Effectiveness at producing sparse representations.  
ii)Competitiveness, with respect to the time required to process 
large images.
The latter is a consequence of the suitability of the
proposed dictionaries for approximating images in partitions of
small blocks.
This feature makes it possible to apply the 
effective greedy selection technique Orthogonal Matching Pursuit,
up to some block size. For blocks exceeding that size
a refinement of the original Matching Pursuit approach is considered. 
The resulting method is termed Self Projected Matching Pursuit,
because is shown to be effective for implementing, via 
Matching Pursuit itself, the optional back-projection 
intermediate steps in that approach. 
\end{abstract}
\maketitle
\section{Introduction}
\label{sec:intro}
A common first step in most image processing techniques is
to map the image onto a transformed space allowing for the reduction 
in the number of points to represent the image, up to some desired
precision. For a significant reduction in the data dimensionality, from say 
$N$ to $K <N$ points, the image is said to be $K$-sparse in
the transformed domain. 
In addition to the many applications that benefit from sparse 
representation of information
\cite{FCR06, MES08, YSS09, WMM10, SMF10}, the emerging theory 
of sampling, called {\em{compressive sensing$/$sampling}},
asserts that sparsity  of a representation may also lead to more economical
data collection \cite{CW08,Rom08, Bar11,XL11,FT12}. 
The relevance of compressive 
sensing within the context of astronomical data is 
discussed in \cite{BSO08,SB09}, where algorithms for signal 
recovery are advanced and illustrated by these types of data.
The usual compressive sensing framework assumes that the signal 
is sparse in an orthogonal basis or incoherent dictionary, because 
most of the recovery proofs have been achieved under those conditions. 
However, recent theoretical results expand the analysis to 
coherent dictionaries \cite{RSV08,CENR10},  because it is often 
the case that an approximation is sparser when elements 
from such a dictionary are used in the decomposition.
Alternatively, as shown in \cite{BRN11,RNBC12},
high sparsity enables exploitation of the redundancy 
in the pixel intensity representation of an image, 
to reduce the image size when encrypted. 
The success of this technique, termed  
Encrypted Image Folding (EIF),  
strongly depends on the sparsity of the image representation. The 
sparser the representation is the smaller the size of the 
folded image.

In this Communication we wish to highlight the significant 
gain in sparsity that may be obtained by releasing the condition of 
incoherence when designing a dictionary for 
sparse representation of astronomical images. The problem we 
address is described as follows: 

{\em{Given an astronomical image, find its sparse decomposition 
as a superposition of elementary components, selected from a large 
redundant set called a {\em{dictionary.}}}}

It is clear that the success of producing
a very sparse representation of an image
depends in a large part on the ability to construct appropriate
dictionaries from  which to select the right elements, frequently
called `atoms'.
Here a mixed dictionary is considered,  which will be shown to be 
suitable for achieving  sparse representations of astronomical images.
A useful dictionary for this purpose should be capable
of sparsely representing two different features;
i)fairly smooth regions (nebulae) of intergalactic media, gases, dust,
etc., and
ii)bright spots (stars).
In order to account for smooth regions  we use
a  Redundant Discrete Cosine (RDC) dictionary. The model of
bright spots and edges is accomplished by the union of   
B-spline dictionaries of different order and support. 
The combination of these two types of dictionaries
provides us with a mixed dictionary yielding a very significant gain
in the sparsity of an astronomical image, 
in relation to the outcomes from the most commomly used transformations  
in image processing; the Discrete Cosine Transform (DCT) and 
Discrete Wavelet Transform (DWT).
Their  convenient distinctive feature is that  
{\em{they are suitable for processing large images by segmenting 
them into small blocks}}. 
The advantage of this property is twofold: a)It entails
storage requirements which are affordable for processing by 
effective pursuit strategies. 
b)The sequential processing of blocks is fast enough to be practical 
and there is also room for 
the possibility of straightforward parallel processing
when those resources become widely available.

The numerical experiments for illustrating the approach
involve large images from the European Southern Observatory (ESO)
\cite{esoweb} and a set of fifty five images captured by the 
Hubble Space Telescope (HST) \cite{hubweb}.   
Considerations are restricted to approximations of high quality
(PSNR of 45 dB or higher).
While the sparsity level strongly depends 
on each particular image, in all the cases is massively higher  
than the sparsity yielded by the  DCT and DWT.
Since the computational time is very competitive,
we confidently conclude that the mixed dictionaries 
under consideration are suitable for achieving highly 
sparse representation of astronomical images. 

The paper is organized as follows: Sec.~\ref{hnla} discusses 
highly nonlinear approximation techniques and 
introduces the discrete B-spline based dictionaries which, 
together with the RDC dictionary, 
form the highly coherent mixed dictionary that provides  the 
basic elements for representing an image.
Matching Pursuit like selection techniques are also discussed 
in this section. In particular, the proposed
Self Projected Matching Pursuit
strategy is established as a possible alternative to  
Orthogonal Matching Pursuit, when the latter 
cannot be implemented due to storage requirements.
Sec.~\ref{experi} illustrates 
the capability and effectiveness of the approach to yield fast 
sparse representation of astronomical images.
The conclusions are presented in Sec.~\ref{conclu}.

\section{Sparse representation by highly nonlinear approximation 
techniques} 
\label{hnla}

We start by introducing some basic notation: $\R$ and 
$\N$ represent the sets of real and natural numbers, respectively.
Boldface letters are used to indicate Euclidean vectors 
or matrices,
whilst standard mathematical fonts indicate components,
e.g., $\vd \in \R^N$ is  a vector of components
$d(i),\, i=1,\ldots,N$  and $\vI \in \R^{\Nx \times \Ny}$
a matrix of elements $I(i,j),\,i=1,\ldots, \Nx, \, j=1,\ldots,\Ny$. 

Let $\D=\{\vd_n \in \R^N\}_{n=1}^M$ be a spanning set for 
an inner product space $\V_N$ of finite dimension $N$ and 
$\vf \in \R^N$ a signal to be approximated 
by an element 
$\vf^K \in \V_K= \Spann \{\vd_{\ell_i}\}_{i=1}^K$, i.e.,
\be
\label{atomd}
\vf^K= \sum_{i=1}^K  c(i) \vd_{\ell_i}, \text{where} \quad K<N.
\ee
When $N=M$ and the spanning set $\D$ is linearly independent it is a basis 
for $\V_N$, otherwise it is a redundant {\em{frame}} \cite{You80}.
In order to advance in the
discussion as to how to select from $\D$ the
$K$-elements $\vd_{\ell_i},\,i=1,\ldots,K$  in \eqref{atomd},
we need to discriminate two different situations:
\begin{itemize}
\item [i)]$M=N$ and  $\{\vd_i\}_{i=1}^M$ is an orthogonal basis for $\V_N$.
\item [ii)]$K < N$ and $\{\vd_i\}_{i=1}^M$ is a non orthogonal 
and not necessarily linearly independent spanning set for $\V_N$. 
\end{itemize}
Case i) leaves rooms for the linear and nonlinear forms of 
selecting the elements $\vd_{\ell_i},\,i=1,\ldots,K$. 
Both types of approximation are easily realized in practice.
A linear  
procedure  determines before hand a fixed order for 
the elements of $\D$  and  uses, say the first $K$ elements,
for the approximation. On the contrary, a nonlinear procedure 
would make the selection dependent on the signal to be approximated. 
For example: it is well known that in order to construct the approximation  
$\vf^K$ of $\vf$, such that $\|\vf^K - \vf\|$ is minimum 
(where $\|\cdot\|$ is the square norm induced by the inner product) 
one should select the elements 
$\vd_{\ell_i},\,i=1,\ldots,K$ corresponding to the $K$ coefficients 
$c(i)= \la \vd_{\ell_i},\vf\ra ,\,i=1,\ldots,K$ 
of largest absolute value. This 
approximation is nonlinear, but the implementation in finite dimension is
 straightforward.
  
On the contrary, case ii) introduces an intractable 
problem. If $K$ is fixed, the choice of the 
$K$ elements $\vd_{\ell_i},\,i=1,\ldots,K$ minimizing 
$\|\vf^K - \vf\|$ involves  a combinatorial problem. 
Moreover, the alternative situation;  the 
one of finding the minimum value of $K$ such that 
$\|\vf^K - \vf\| < \rho$, for a given tolerance $\rho$, 
is also intractable. 
This is the reason why this type of approximation is said to be 
{\em{highly non linear}}, and in practice is addressed in some 
suboptimal manner. Rather than looking for the 
sparsest solution (minimum value of $K$) one looks 
for a `sparse enough solution'. This means that
the number of $K$-terms in \eqref{atomd} is `small enough' for 
the representation to be convenient in the particular context.

Usually highly non linear approximations of a signal 
$\vf$ using a dictionary $\D=\{\vd_i\}_{i=1}^M$ are 
realized by: 
\begin{itemize}
\item [a)] Expressing $\vf^K$ as $\sum_{i=1}^M c(i) \vd_i$ and 
finding $K$-nonzero coefficients by minimization of the 1-norm  
$\|\vc\|_1= \sum_{i=1}^M |c(i)|$ \cite{CDS98}.
\item [b)] Using  a greedy pursuit strategy for 
stepwise selection of the $K$ normalized to unity
elements $\vd_{\ell_i}\,i=1,\ldots,K$, 
called {\em{atoms}}, for producing
the approximation $\vf^K= \sum_{i=1}^K c(i) \vd_{\ell_i}$, 
which is termed {\em{atomic decomposition}}.
\end{itemize}
We restrict considerations to greedy pursuit algorithms
because, for the highly coherent dictionaries we are considering,
are more effective and  
faster than those based on minimization of the 1-norm. 
\subsection{Matching Pursuit based selection techniques}
\label{mp}
The greedy selection strategy Matching Pursuit (MP) 
was introduced with this name in the context of 
signal processing by S. Mallat and  Z. Zhang \cite{MZ92}. 
Previously it had appeared as a regression technique in 
the statistical literature, where the convergence property 
was established \cite{Jon87}. The implementation is very simple and 
evolves by successive approximations as follows.

Let $\vRe^{k}$ be the $k$-th order residue, 
$\vRe^{k}= \vf -\vf^k$, and $\ell_{k}$ the index for
which the corresponding dictionary atom $\vd_{\ell_{k}}$ yields 
a maximal value of $|\la \vd_{i} , \vRe^{k} \ra|,\, i = 1,\ldots M$.
Starting with an initial approximation
$\vf^1= 0$ and $\vRe^{1} = \vf$ the algorithm 
iterates by sub-decomposing the $k$-th order residue into
\be
\vRe^{k} =
\la \vd_{n}, \vRe^{k} \ra \vd_{n} + \vRe^{k+1}, \quad  n = 1,\ldots, M,
\label{tech:1}
\ee
which defines the residue at order $k+1$.
Since $\vRe^{k+1}$ given in \eqref{tech:1} is orthogonal to all $\vd_{n}, 
\, n = 1,\ldots, M$, it is true that 
\be
\|\vRe^{k}\|^{2} = |\la \vd_{n}, \vRe^{k} \ra|^{2} + \|\vRe^{k+1}\|^{2}, 
\quad  n = 1,\ldots, M.
\label{tech:2}
\ee
Hence, the dictionary atom $\vd_{\ell_k}$ yielding a maximal value  
of $|\la \vRe^{k}, \vd_{n} \ra|$ minimizes $\| \vRe^{k+1}\|^{2}$.

From \eqref{tech:1} it follows that at each iteration $k$ the MP 
algorithm results in an intermediate representation of the form:
\be
\vf= \vf^k + \vRe^{k+1}
\label{tech:3}
\ee
with
\be
\vf^{k}= \sum_{n=1}^{k} \la \vd_{\ell_{n}}, \vRe^{n} \ra \vd_{\ell_{n}}.
\label{tech:4}
\ee
It was first proved in \cite{Jon87} 
that in the limit $k \rightarrow \infty$ the sequence  $\vf^{k}$
converges to $\hat{P}_{\V_{M}}\vf $, 
the orthogonal projection of $\vf$ onto 
$\V_{M}= \Spann\{\vd_{{n}}\}_{n=1}^M$      
(the proof is translated to the MP context in \cite{MZ92}).
Nevertheless, if the algorithm is stopped at the $k$th-iteration, 
$\vf^{k}$ recovers an approximation of $\vf$ with an 
error equal to the norm of the residual $\vRe^{k+1}$ which, if 
the selected atoms are not 
orthogonal, will not be orthogonal to the subspace 
 they span. An additional drawback of the MP approach is that 
 the selected atoms may not be linearly independent.

A refinement to MP, which does yield an 
orthogonal projection approximation at each step, has been termed 
Orthogonal Matching Pursuit (OMP)
\cite{PRK93}. 
In addition to selecting only linearly independent atoms, 
the OMP approach improves upon MP numerical convergence rate 
and therefore amounts to be, usually,
a better approximation of a signal after a finite number of iterations. 
OMP provides a decomposition of the signal as given by:
\begin{equation}
\vf^k = \sum_{n=1}^{k} c^k(n) \vd_{\ell_{n}} + \tilde{\vRe}^{k},
\label{tech:5}
\end{equation}
where the coefficients $c^k(n)$ are computed in such a way that 
it is true that
$$\sum_{n=1}^{k} c^k(n) \vd_{\ell_{n}}= \hat{P}_{\V_{k}}\vf,\quad{\text{with}}
\quad \V_{k}= \Spann\{\vd_{\ell_{n}}\}_{n=1}^k.$$ Thus, 
OMP yields the unique element $\vf^k \in \V_{k}$ minimizing 
$\|\vf^k -\vf\|$.
The superscript of $c^k(n)$ in \eqref{tech:5} indicates 
the dependence of these quantities on the iteration step $k$. 

The OMP approach is effective for processing signals up to 
some dimensionality.
It may become  prohibitive, because of its storage requirements, when the
signal dimension exceed  some value. In this respect, 
 MP has the advantage of being  suitable  for processing 
  very large dimensional signals and, 
for 2D images, it fully exploits the separability  property 
of dictionaries.
Since our mixed dictionaries are adequate for block processing,
in general the OMP approach is an appropriate technique. However,  
 one of the aims of the present effort is 
  to study, in a standard personal computer, 
the dependence of the sparsity of a representation with respect to 
the block size of an image partition. For this purpose, 
we are forced to overcome  storage requirements of the standard  
OMP implementations.
The goal is achieved by
applying the refinement to the MP method proposed in the next section.
\subsection{Self Projected Matching Pursuit}
\label{SPMP}
The seminal paper \cite{MZ92} discusses a possible 
 improvement of the MP approximation by
means of back-projection steps, which stands 
for computing the orthogonal projection
of the MP approximation. The authors suggest this could be 
done by the conjugate gradient method. Unfortunatelly 
 the processing time of that method is not affordable 
 in practice for large dimensional problems, and specially 
 with highly correlated dictionaries.
 Thus, the question we have tried to answer is:

{\em{Since the MP approach converges asymptotically 
to the orthogonal projection onto
the span of the selected atoms,  would it be affective
to use MP itself to compute the back-projection steps?}}

Of course there is an increment of step wise
complexity but, as the example presented here 
illustrates,  on the whole  the refinement may 
perform  better and faster. 

The resulting method, that we have termed
Self Projected Matching Pursuit (SPMP) evolves as follows.
Given a dictionary $\D=\{\vd_n\}_{n=1}^M$ and a signal $\vf$,
set $S=\{\emptyset\}$ and $\vR=\vf$. Assuming that the required 
projection step is of length $p$, implement the  algorithm
below. 
\begin{itemize}
\item[i)]
Apply MP up to step $p$ selecting atoms from 
dictionary $\D=\{\vd_n\}_{n=1}^M$. Assuming 
that the distinct selected atoms  are $\{\vd_{\ell_{n}}\}_{n=1}^{k}$ 
assign $S \leftarrow S \cup \{\vd_{\ell_{n}}\}_{n=1}^{k}$. 
Set $K$  equal to the cardinality of the updated $S$. Let us denote as
$\vf^K$ the approximation of $\vf$ so far and as $\vRe^K$ the 
residue $\vRe^K= \vf -\vf^K$.
\item[ii)]
Approximate $\vRe^K$ using only the selected set
$S$ as the dictionary, which guarantees the  asymptotic
convergence to the 
approximation $\op_{\S}{\vRe^K}$ of $\vRe^K$,  where $\S= \Spann \, S$,
and a residue $\vRe^\perp= {\vRe^K} - \op_{\S}\vRe^K$ having no 
component in $\S$.
\item[iii)] Set $\vRe \leftarrow \vRe^\perp$ and $\vf^K \leftarrow  \vf^K + \op_{\S}\vRe $ and repeat steps 
i) and  ii) until, for a required  $\rho$, the 
condition $\|\vRe\| < \rho$ is reached. 
\end{itemize}
For $p=1$ the above refinement gives, asymptotically, 
the orthogonal projection approximation at each iteration,
thereby reproducing the results of OMP. As illustrated by the 
example below, significant improvement upon the original 
MP approach may be achieved for values of $p$ larger than one.

{\bf{Example.}} This numerical example is a hard test for MP. 
Consider the Redundant Discrete Cosine (RDC) dictionary
$\D_1$  given by:
\ben
\label{RDC}
\D_1=\{\vv_i;\,v_i(j)= w_i\cos(\frac{\pi(2j-1)(i-1)}{2M}),\,j=1,\ldots,N
\}_{i=1}^M,
\een
with $w_i,\,i=1,\ldots,M$ normalization factors.
For $M=N$ this set is
a Discrete Cosine (DC) orthonormal basis for the
Euclidean space $\R^N$.
For $M=2zN$, with $z \in \N$, the set is a RDC dictionary
with redundancy $2z$, which will be fixed equal to 2.
\begin{figure}
\begin{center}
\includegraphics[width=9cm]{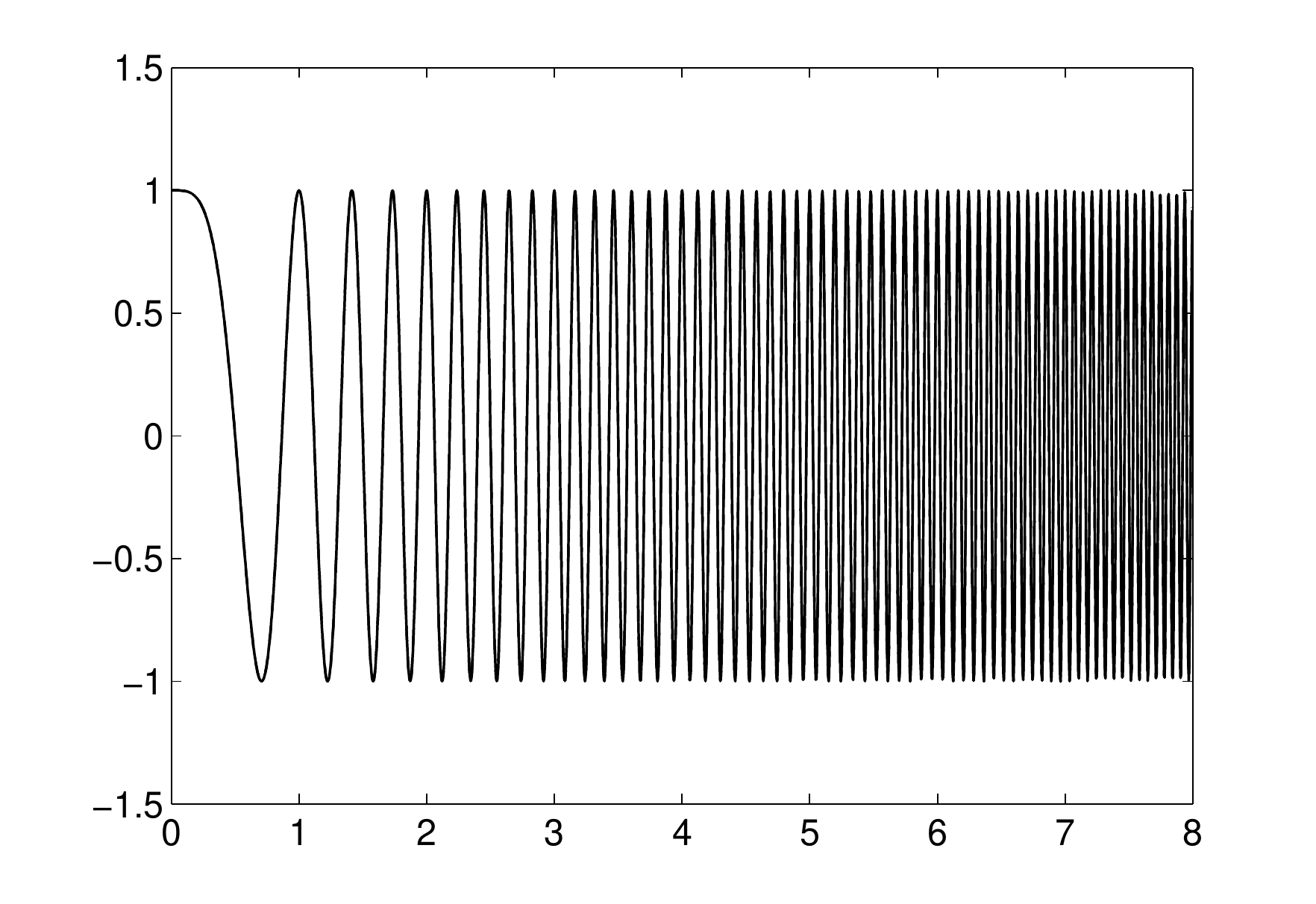}
\end{center}
\vspace{-0.7cm}
\caption{\small{Chirp signal approximated up to error $\rho=0.001 \| \vf \|$  
by i) $K=683$ orthogonal 
DC components taken from \eqref{RDC} with $N=M=2000$.  ii) $K=286$ 
atoms taken from  \eqref{RDC} with $M=2N=4000$ using OMP or 
$K= 1638$  using MP . iii) 
$K=300$ atoms taken from \eqref{RDC} with $M=2N=4000$ using SPMP
with $p=10$ or $K=286$ with $p=3$.}}
\label{chirp}
\end{figure}
To represent the chirp signal $\cos( 2\pi t^2)$ depicted in 
Fig.~1 we take an equidistant partition of the interval 
$[0,8]$ consisting of $N=2000$ points and sample the
chirp at those points $f(i),\,i=1,\ldots, N$. The  
aim is to find an approximation of these  points
up to precision $\rho= 0.001 \|\vf\|$.
Considering $M=N=2000$ in the above definition 
of $\D_1$ we have an orthonormal basis and therefore both 
MP and OMP methods give the 
sparsest decomposition of the signal in orthogonal DC  
components. For an 
approximation to the given precision (coinciding visually with 
the theoretical chirp in Fig.~1) it is necessary to 
use $K=683$ orthogonal elements from \eqref{RDC}. 
Now, setting $M=2N=4000$ the 
dictionary $\D_1$ is no longer an orthogonal basis but a 
redundant {\em{tight frame}} \cite{You80} and the algorithms 
MP and OMP produce very different decompositions. While 
OMP improves the sparsity of the representation requiring
only $K=286$ components, MP needs $K=1638$ different atoms, i.e. 
significantly more  than with the orthonormal basis.
The reason for the poor performance of MP is that 
in the redundant dictionary the atoms are  highly correlated and 
the method is picking linearly dependent atoms, something that cannot occur
 with OMP. However, when applying the proposed 
refinement SPMP with projection step $p=10$ the number of 
required components 
drops to $K=300$. For $p=3$ the number of required components is 
that of OMP, i.e. $K=286$. While in this  example 
there is no need for the SPMP approach,  because the  
already established algorithm OMP performs the decomposition faster,
the result illustrates the fact that SPMP can provide  an 
effective alternative to OMP when, as is  the case 
with 2D images, OMP becomes  
slow or its storage demands  cannot be met. Further details for
the 2D implementation of SPMP will be discussed in Sec.~\ref{omp2d}.  
Before that, we shall introduce the proposed mixed dictionaries for 
representing astronomical images.
\subsubsection{Building mixed dictionaries for sparse 
representation of astronomical images}
\label{dic}
Assume that the $K$-sparse representation of a given 
image $\vI \in \R^{N_x \times N_y},\,$ 
$N_x, N_y \in \N$, is represented as
\be
\label{atomic}
\vI^{K}= \sum_{i=1}^K c^K(i) \vD_{\ell_i},
\ee
where the elements $\vD_{\ell_i}\in \R^{N_x \times N_y},
\,i=1,\ldots,K$ in \eqref{atomic},  
are to be selected from a dictionary $\D=\{D_i\}_{i=1}^{M}$ 
which is obtained as the Kronecker product  $\D=\DX \otimes \DY$
of the dictionaries $\D^x =\{\vd^x_n \in \R^{\Nx}\}_{n=1}^{\Mx}$
and $\D^y =\{\vd^y_m \in \R^{\Ny}\}_{m=1}^{\My}$.
In this section we discuss a particular 
dictionary $\D$, which will be shown to be adequate for 
sparse representation of astronomical images.
\subsection*{Redundant Discrete Cosine (RDC) Dictionary}
As already mentioned in order to sparsely represent the fairly smooth
regions of the images being considered, one of the components of
the proposed mixed dictionary
is chosen to be a RDC dictionary $\D_1$ introduced in the last section, 
fixing $M=2N$ so as to have a RDC dictionary with redundancy two. 
\subsection*{Redundant Discrete B-Spline (RDBS) based dictionaries}
The other component of the proposed mixed dictionary, 
which allows for the representation of bright spots and edges,
is inspired by a general result holding for continuous spline 
spaces. Namely, that {\em{spline spaces on a compact interval
can be spanned by dictionaries of B-splines
of broader support than the corresponding B-spline basis functions 
\cite{AR05,RNX10}.}} 
This may result in a very considerable gain in sparsity for
functions well approximated in spline spaces. 
Here we consider equally spaced knots so that the corresponding B-splines 
are called cardinal. All the cardinal
B-splines of order $m$ can be obtained from
one cardinal B-spline $B(x)$ associated with the uniform simple
knot sequence $\Del= 0,1,\dots,m$.
Such a function is given as \cite{Boo78}
\begin{equation}
B_m(x)=\frac{1}{m!}\sum_{i=0}^m(-1)^i\binom{m}{i}(x-i)^{m-1}_+,
\end{equation}
where $(x-i)^{m-1}_+$ is equal to $(x-i)^{m-1}$ if $x-i>0$ and 0 otherwise.
We shall consider only B-Splines of order $m=2$ and $m=4$ and include 
associated derivatives. 
For $m=2$ the corresponding space is the
space of piece wise linear functions and can be spanned by
a linear B-spline basis, or dictionaries of broader support, 
arising by translating a prototype `hat' function. 
Equivalently, the cubic spline space corresponding to $m=4$
is spanned by the usual cubic B-spline basis, or dictionaries of
cubic B-spline functions of broader support.
Details on how to build B-spline
dictionaries are given in \cite{AR05,RNX10}. 
The numerical construction of the cases $m=2$ and 
$m=4$ considered here 
is very simple and arises by translations of the prototype functions 
given below:

\begin{subnumcases}
{\label{B2} B^l_2(x)=}
\frac{x}{l} & \mbox{if}\quad $0 \leq x <  l$\\
2-\frac{x}{l} & \mbox{if}\quad  $l\leq x <2l$ \\
 0 &  \mbox{otherwise}.
\end{subnumcases}

\begin{subnumcases} 
{\label{B4} B^l_4(x)=}
\frac{x^3}{6 l^3} & \mbox{if} \quad  $0  \leq  x<l$\\
-\frac{x^3}{2l^3} + 2\frac{x^2}{l^2} -2 \frac{x}{l} + \frac{2}{3} & \mbox{if} \quad $l \leq  x< 2l$\\
\frac{x^3}{2l^3}- 4\frac{x^2}{l^2} + 10\frac{x}{l} -\frac{22}{3} & \mbox{if} \quad $2l \leq  x< 3l$\\
-\frac{x^3}{6l^3} + 2\frac{x^2}{l^2} -8\frac{x}{l} +\frac{32}{3} & \mbox{if} \quad $3l \leq  x< 4l$\\
0 & \mbox{otherwise}.
\end{subnumcases}
The B-spline basis for the cardinal spline space 
corresponding to $m=2$ is constructed 
by considering $l=1$ in 
\eqref{B2} and translating the prototype every knot. 
Dictionaries for the identical space of functions of broader 
support arise by setting  $l \in \N$ 
 in  order to fix the desired support. 
The B-spline basis for the cubic cardinal spline  space,
corresponding to $m=4$, requires to set $l=1$ in 
\eqref{B4} and translate the concomitant prototype. 
Dictionaries are obtained  by  taking larger values of $l$. 

\begin{figure}
\begin{center}
\includegraphics[width=7cm]{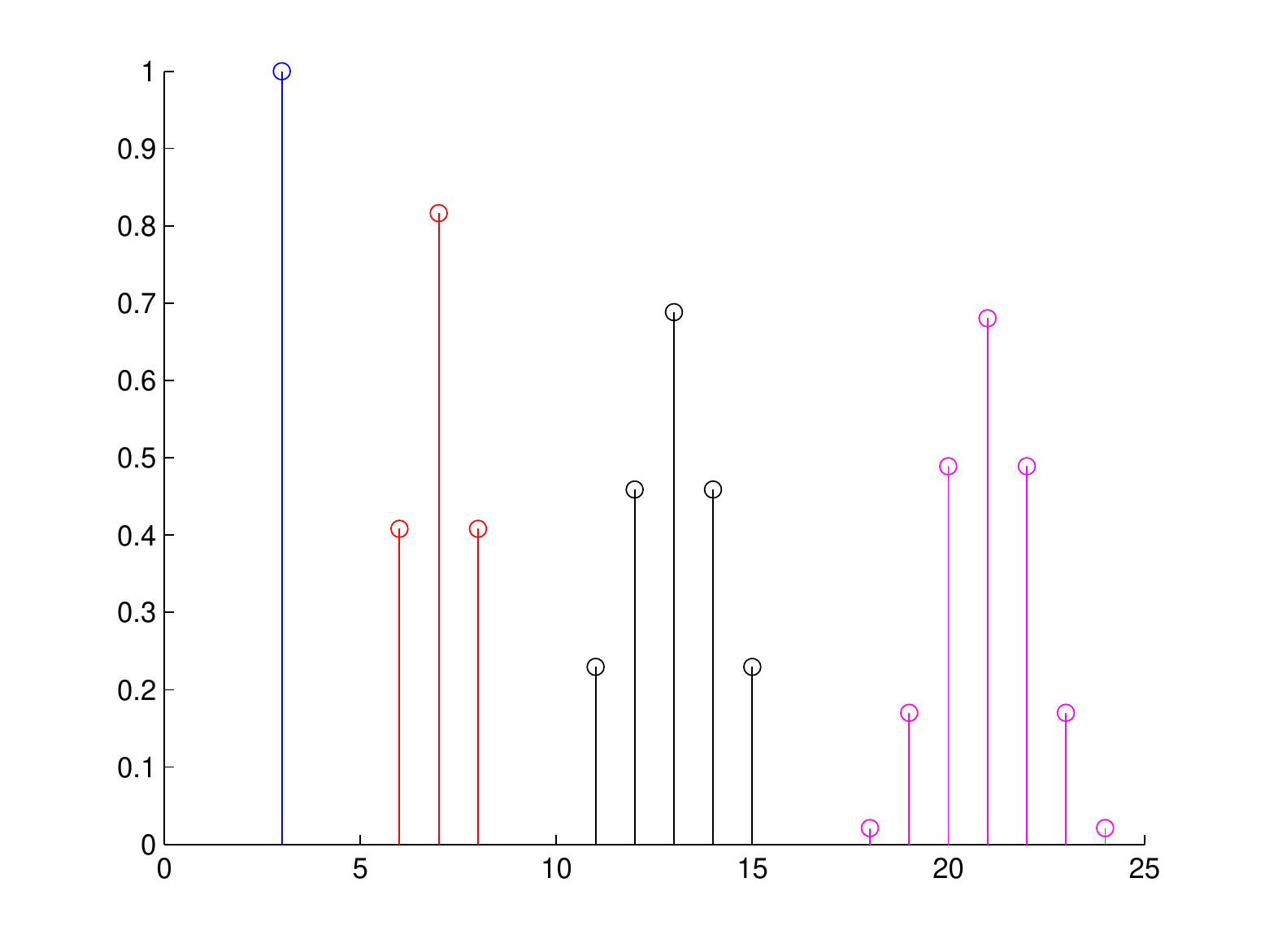}
\includegraphics[width=7cm]{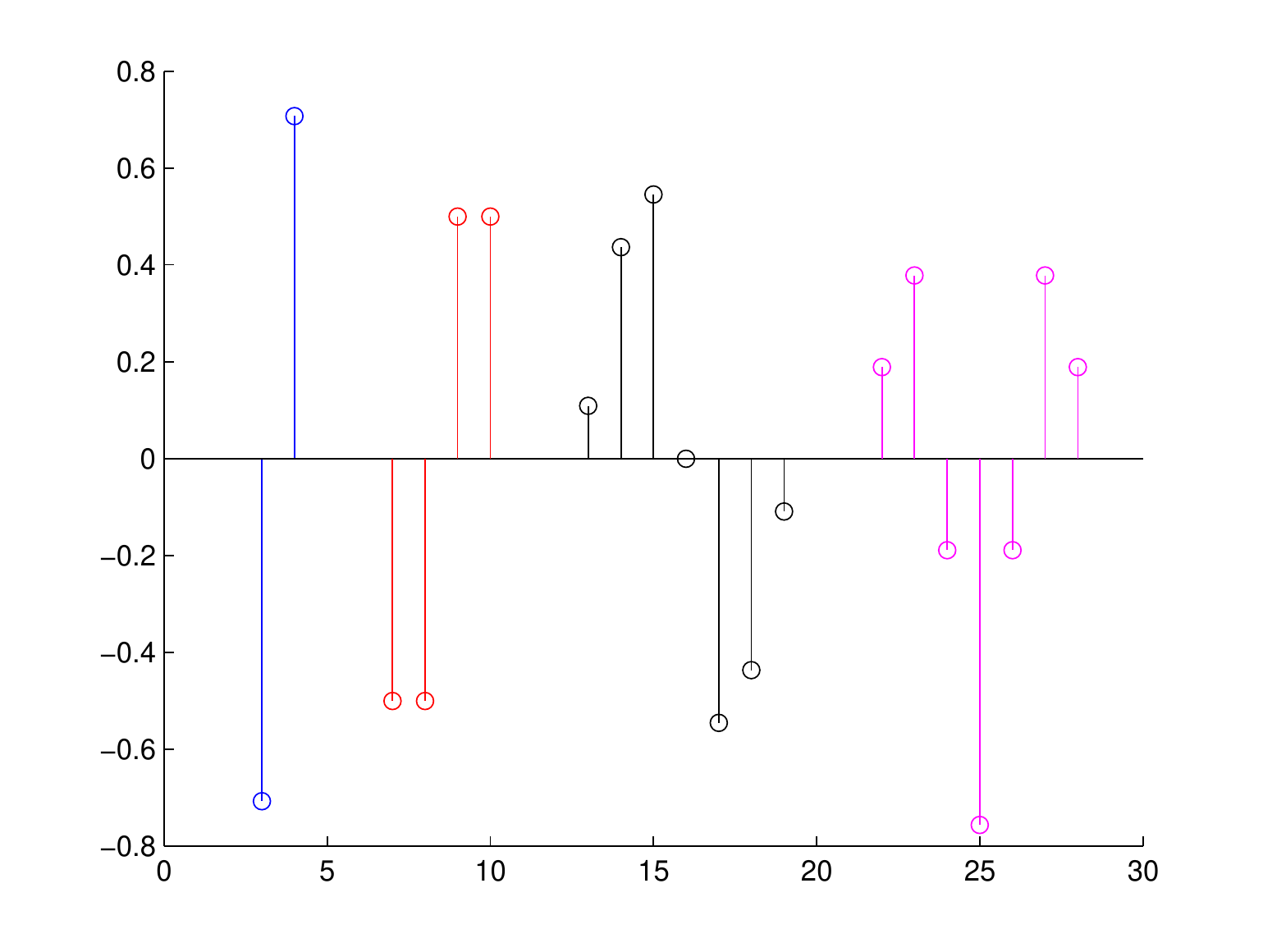}
\end{center}
\vspace{-0.75cm}
\caption{\small{Prototype atoms as defined in \eqref{Ym2} and
\eqref{Ym4}. The RDBS component of the dictionary is constructed by 
translation or these atoms, applying the cut off approach at the 
 boundaries.}}
\label{splinesa}
\end{figure}

As discussed below, derivatives of the above functions 
also provide  suitable 
prototypes to achieve higher
levels of sparsity in the representation of a signal.  
Now, for constructing dictionaries for digital  
image processing  we need to 
\begin{itemize}
\item[a)]Discretize the functions to obtain adequate Euclidean 
vectors.
\item[b)]Restrict the functions to intervals which allows 
images to be approximated in small blocks.
\end{itemize}
We carry out the discretization by taking the value of a
prototype function only at the knots (c.f. small circles
in graphs Fig.~\ref{splinesa}) and translating the 
prototype one sampling point at each translation step. 
At the boundaries we apply the `cut off' approach and keep 
all the vectors whose support has nonzero intersection with the 
interval being considered. 
\begin{remark}
It is worth mentioning that by the proposed discretization 
the hat B-spline basis for the corresponding 
interval becomes the standard Euclidean basis.
By discretizing the hats of broader support 
the samples preserve the hat shape.
\end{remark}
Obviously for a finite dimension Euclidean space one can construct 
arbitrary dictionaries.
In particular, redundant B-spline based dictionaries with 
prototypes of different support and shapes arising from
the functions \eqref{B2} 
and \eqref{B4} and their corresponding derivatives.

Indicating as ${\rm{d}}^1B^l_m(x)$ the derivative of $B^l_m(x)$ and
as ${\rm{d}}^2B^l_m(x)$ its second derivative, in 
addition to linear and cubic B-splines  we  shall consider
the additional prototypes
${\rm{d}}^1B^l_2(x),\, {\rm{d}}^1B^l_4(x)$ and ${\rm{d}}^2B^l_4(x)$.
Vectors of different support may be included by 
merging those dictionaries. For our 
experiments we construct the Redundant  Discreet B-Spline (RDBS) 
based dictionaries as follows: 
$$\D_s=\{b_iY^s_m(j-i)|N; j=1,\ldots,L \}_{i=1}^{M_s},\quad
m=2,4,\,s=2,\ldots,9,$$
where the notation $Y_m(j-i)|N$ indicates the restriction to be
an array of size $N$
and $b_i, \,i=1,\ldots,M_s$ are normalization constants. 
The arrays $Y^s_2,\,s=2,3,4,\, Y^7_4$, shown consecutively in the 
left graph of Fig.~\ref{splinesa}, and $Y^s_2,\, s=5,6,\  
Y^s_4, s=8,9$ shown consecutively in the right graph of the 
same figure, are defined as follows:
\begin{subnumcases}
{\label{Ym2} Y^s_2=}
B^l_2,   \,l=1,2,3 &\mbox{for} \quad s=2,3,4\, \mbox{(respectively)}\\
{\rm{d}}^1B^l_2,\,l=2,3   &\mbox{for} \quad s=5,6\,   \mbox{(respectively).}
\end{subnumcases}

\begin{subnumcases}
{\label{Ym4} Y^s_4=}
B^2_4    &\mbox{for}\quad s=7 \\
{\rm{d}}^1 B^2_4 &\mbox{for}\quad s=8\\
{\rm{d}}^2 B^2_4 &\mbox{for}\quad s=9.
\end{subnumcases}

The cut off approach applied to the boundaries implies that 
the numbers $M_s$ of total atoms in the $s$th-dictionary
varies according to the atom's support. 

Taking $N=N_x$, an unidimensional mixed dictionary, $\DX$, 
results by joining 
dictionary $\D_1$ (c.f. \eqref{RDC})  and the above 
defined RDBS ones, i.e.
$\DX= \cup_{s=1}^{9} \D_s,$. Taking $N=N_y$ an equivalent 
mixed dictionary, $\DY$, is obtained. The mixed dictionary $\D$
for $\R^{N_x \times N_y}$ is 
the Kronecker product  $\D=\DX \otimes \DY$. 
However, as discussed below,  this 
 2D dictionary does not need to be constructed.
This advantage represents a huge save in memory requirements.

\subsection{2D implementation of the selection strategies with  separable 
dictionaries}
\label{omp2d}
Given an image $\vI \in \R^{\Nxy}$ and two 1D dictionaries
$\D^x =\{\vd^x_n \in \R^{\Nx}\}_{n=1}^{\Mx}$ and
$\D^y =\{\vd^y_m \in \R^{\Ny}\}_{m=1}^{\My}$
the greedy procedure
OMP2D for approximating $\vI$  with atoms taken from
$\D^x$ and $\D^y$ iterates as follows.

On setting $\vRe^0=\vI$ at iteration $k+1$ the algorithm selects the atoms
$\vd^x_{\ell^x_{k+1}} \in \D^x$ and $\vd^y_{\ell^y_{k+1}} \in \D^y$
that maximize the absolute value of the Frobenius inner products
$\la \vd^x_n ,\vRe^{k} \vd^y_m \ra_\F,\, n=1,\ldots,\Mx,\,m=1,\ldots,\My,
$ i.e.,
\be
\begin{split}
{\llx_{k+1},\lly_{k+1}}&= \operatorname*{arg\,max}_{\substack{n=1,\ldots,\Mx\\
m=1,\ldots,\My}} |\sum_{\substack{i=1\\j=1}}^{{\Nx,\Ny}}
d^x_{n}(i) R^k(i,j) d^y_{m}(j)|\\
\text{with}\\
R^{k}(i,j)& = I(i,j) - \sum_{n=1}^{k} c^k(n)
d^x_{\ell^x_n} (i) d^y_{\ell^y_n}(j).
\end{split}
\label{omp}
\ee
The coefficients $c^k(n),\,n=1,\ldots,k$ in the above expansion
are such that $\|\vRe^{k}\|_\F$ is minimum, where $\|\cdot \|_\F$ is 
the Frobenius norm. 
This is ensured by 
requesting that $\vRe^{k}= \vI- \op_{\V_k} \vI$,
where $\op_{\V_k}$ is the orthogonal
projection operator
onto $\V_k=\Spann\{\vd^x_{\ell^x_n} \otimes \vd^y_{\ell^y_n} \}_{n=1}^k$.
A straightforward generalization of the implementation discussed 
in \cite{RNL02,ARN06} for the 1D case provides us with the
representation of $\hat{P}_{\V_k} \vI$ as given by
\be
\label{oproj}
\hat{P}_{\V_k} \vI = \sum_{n=1}^k \vA_n \la \vB_n^k, \vI \ra_F =
\sum_{n=1}^k  c^k(n) \vA_n,
\ee
where each $\vA_n \in \R^{\Nxy}$ is an array with the
selected atoms $\vA_n= \vd^x_{\ell^x_n} \otimes \vd^y_{\ell^y_n}$
and  $\vB_n^k,\,n=1,\ldots,k$  are the
concomitant reciprocal matrices,
which are the unique elements
of $\R^{\Nxy}$  satisfying the conditions:
\begin{itemize}
\item
[i)]$\la \vA_n, \vB_m^k \ra_\F=\delta_{n,m}= \begin{cases}
1 & \mbox{if}\, n=m\\
0 & \mbox{if}\, n\neq m.
\end{cases}$
\item
[ii)]${\V_k}= \Spann\{\vB_n^k\}_{n=1}^k.$
\end{itemize}
Such matrices can be adaptively constructed through the recursion formula:
\be
\begin{split}
\label{BW}
\vB_n^{k+1}&= \vB_n^k - \vB_{k+1}^{k+1}\la \vA_{k+1}, \vB_n^k \ra_\F,\quad n=1,\ldots,k\\
\text{where}\\
\vB_{k+1}^{k+1}&= \vW_{k+1}/\| \vW_{k+1}\|_\F^2,\,\,
\text{with}\,\,
\vW_1=\vA_1 \,\, \text{and} \,\,
\vW_{k+1}= \vA_{k+1} - \sum_{n=1}^k \frac{\vW_n}
{\|\vW_n\|_\F^2} \la \vW_n, \vA_{k+1}\ra_\F.
\end{split}
\ee
For numerical accuracy in the construction 
of the set
$\vW_{n},\,n=1,\ldots,k+1$ at
least one re-orthogonalization step
is usually needed. It
implies that one needs to recalculate these matrices as
\be
\label{GS}
\vW_{k+1}= \vW_{k+1}- \sum_{n=1}^k \frac{\vW_n}{\|\vW_n\|_\F^2}
\la \vW_n , \vW_{k+1}\ra_\F.
\ee
The coefficients in \eqref{oproj} are obtained from the
inner products
$c^k(n)= \la \vB_n^k, \vI \ra_\F,\, n=1,\ldots,k$. The
algorithm iterates up to step, say $K$, for which, for a given $\rho$,
the stopping criterion $\|\vI - \vI^K\|_\F < \rho$ is met.
The MATLAB function OMP2D, and corresponding MEX file in C{\small{++}}
for faster implementation of the identical function, are
available at \cite{webpage2}. 

Up to some block-size OMP2D is very effective. It takes advantage 
of the separability property of the dictionary, 
except for the construction of 
the required matrices  $\vB_n^k ,\, n=1,\ldots,k$. Unfortunately, for 
blocks larger than a certain size  
the concomitant storage demands are
not available on a standard personal computer, or the 
computations became very slow and the above implementation of OMP2D 
is no longer affective.
As already mentioned,
in order to avoid the storage and computation of matrices 
$\vB_n^k ,\, n=1,\ldots,k$,  
we propose the SPMP method. Algorithms 1, 2, and 3 
outline its implementation in 2D, which we term SPMP2D. 
\newcounter{myalg}
\begin{algorithm}[!ht]
\refstepcounter{myalg}
\begin{algorithmic}
\STATE{\bf{Input:}} Image $\vI \in \R^{\Nx\times \Ny}$.
Dictionaries $\D^x=\{\vd^x_n \in \R^{\Nx}\}_{n=1}^{\Mx}$ and 
$\D^y=\{\vd^y_n \in \R^{\Ny}\}_{n=1}^{\My}$.\\
Approximation error $\rho>0$ and tolerance $\epsilon>0$ for
the numerical realization of the projection.
Length of projection step $p$.\\

\STATE{\bf{Output:}} Approximated image $\IA \in  \R^{\Nx \times \Ny}$.
Coefficients in the atomic decomposition $\vc \in \R^k$. 
Ordered pair of indices labeling the selected atoms 
$\Gamma=\{(\ell^x_n, \ell^y_n)\}_{n=1}^K$.\\

\STATE \COMMENT{{ Initialization}} 

\STATE{\rm{Set}}$\,\,\Gamma\,=\,\{\emptyset\},\quad \IA\,=\,0,\quad
\vR \,=\,\vI\,\quad k\,=\,0, \quad \EA= 2\rho$ 

\STATE{\rm{Set}}$\,\,C(i,j)\,=\,0,\, i\,=\,1,\ldots,\Mx,\, j\,=\,1,\ldots,\My.$ 

\STATE\COMMENT{{ Begin the algorithm}}

\WHILE {$\EA > \rho$} 

\STATE Apply Algorithm~\ref{mpsteps} 
\COMMENT{{{p-MP-iterations}}} to obtain:

\STATE $\Gamma = \{(\ell^x_n, \ell^y_n)\}_{n=1}^k, 
\IA \in \R^{\Nx\times \Ny},\, \vR= \vI - \IA \in \R^{\Nx\times \Ny}$  
and $\vC \in \R^{\Mx \times \My}$

\STATE\COMMENT{{ {Collect nonzero coefficients in $\vc \in \R^k$}}}

\FOR {$n=1:k$}
 \STATE   $c(n)=C(\ell^x_{n}, \ell^y_{n})$
\ENDFOR
\STATE Apply Algorithm ~\ref{spmp} \COMMENT {{ {Improve approximation by orthogonal projection via MP}} } to  update
$\vc \in \R^k$  so that $\IA \leftarrow \IA+  \hat{P}_{\V_k} \vR$,\, 
$\vR \leftarrow 
\vR - \hat{P}_{\V_k} \vR$
\COMMENT {{{where $\V_k=\Spann\{\vd^x_{\ell^x_n} \otimes \vd^y_{\ell^y_n} \}_{n=1}^k$}}}
\FOR {$n=1:k$}
\STATE    $C(\ell^x_{n}, \ell^y_{n})=c(n)$\COMMENT{{{Update 
    of matrix with coefficients}}}
\ENDFOR
\STATE $\EA \leftarrow \| \vI -  \IA \|_F$
\ENDWHILE
\end{algorithmic}
\caption{\small{Implementation of the proposed SPMP2D method to
approximate an image.}}
\label{algo}
\end{algorithm}
\begin{algorithm}[!ht]
\refstepcounter{myalg}
\begin{algorithmic}
\FOR {$t=1:p$} 
\FOR {$n=1:\Mx$ and $m=1:\My$}  

\STATE $G(n,m)=\sum_{\substack{i=1\\j=1}}^{\Nx,\Ny} \dx_{n}(i) R(i,j) \dy_{m}(j)$

\ENDFOR


\STATE ${q^x,q^y}= \arg\max\limits_{n,m=1,\ldots,M}|G(n,m)|$ \COMMENT{{{MP selection of atoms}}}

\STATE\COMMENT{{{Store coefficients as a matrix adding contributions 
from repeated atoms}}}

\STATE $C(q^x, q^y)= C(q^x, q^y)+
    G(q^x ,q^y)$

\STATE\COMMENT{{{Update of approximation and residual}}} 

\FOR {$i=1 :\Nx$  and $j=1:\Ny$}

\STATE $\Delta= C(q^x, q^y) \dx_{q^x}(i)
\dy_{q^y}(j)$

\STATE $R(i,j)=R(i,j)- \Delta$

\STATE $I{_\text{Approx}}(i,j)= I_{\text{Approx}}(i,j)+ \Delta$
\ENDFOR
\IF {$(q^x, q^y) \notin \Gamma$}
\STATE $k \leftarrow k+1, \quad (\ell^x_{k}, \ell^y_{k}) \leftarrow (q^x, q^y),\quad 
\Gamma \leftarrow \Gamma  \cup (\ell^x_{k}, \ell^y_{k})$
\COMMENT{{Update set of indices}}
\ENDIF
\ENDFOR
\end{algorithmic}
\caption{$p$-plain MP iterations for atoms selection  and collection 
of contributions of repeated atoms}
\label{mpsteps}
\end{algorithm}
\begin{algorithm}[!ht]
\refstepcounter{myalg}
\begin{algorithmic}
\STATE {\rm Set} $\,\,\EP= 2\epsilon$

\WHILE {$\EP > \epsilon$}

\FOR {$n=1:k$}  

\STATE	$g(n)=\sum_{\substack{i=1\\j=1}}^{\Nx,\Ny}
	d^x_{\ell^x_{n}}(i) R(i,j)d^y_{\ell^y_{n}}(j)$

\ENDFOR 

	\STATE ${q}=\arg\max\limits_{n=1,\ldots,k} | g(n)| $

	\STATE $c(q)=c(q)+g(q)$ \COMMENT{Update coefficients}
	\STATE\COMMENT{Update approximation and residual}
\FOR {$i=1:\Nx$ and $j=1:\Ny$}
\STATE $\Re(i,j)= \Re(i,j)-  \dx_{\ell^x_{q}}(i) \dy_ {\ell^y_{q}}(j)g(q)$
	\STATE $\I{_\text{Approx}}(i,j)= \I{_\text{Approx}}(i,j) + \dx_{\ell^x_{q}}(i) \dy_ {\ell^y_{q}}(j)g(q) $
\ENDFOR
      \STATE  $\EP \leftarrow |g(q)|$
\ENDWHILE
\end{algorithmic}
\caption{\small{Orthogonal projection onto
 $\V_k=\Spann\{\vd^x_{\ell^x_n} \otimes \vd^y_{\ell^y_n} \}_{n=1}^k$ via MP}}
 \label{spmp}
\end{algorithm}
Putting aside the complexity for the selection process, which is the same 
for both approaches, the complexity order 
for the procedure of including one more 
term in the approximation is
\begin{itemize} 
\item 
O$(k\Nx\Ny)$ for OMP2D 
\item
O$(k \kappa\Nx \Ny)$ for SPMP2D, where $\kappa$ indicates
the number of iterations 
to improve the MP2D approximation by self projections.
\end{itemize}
The number $\kappa$ is expected to depend  on the 
correlation of the selected atoms. 
For the dictionaries and the class of images we are considering we 
can assert that  $\kappa$ is a small number. 
When this relation is fulfilled the complexity 
of both approaches are of equivalent order.
However, as illustrated in the next section, the storage
requirements of OMP2D 
slow the processing significantly when the size of the blocks 
partitioning the images increases 
beyond some value. In such situations SPMP2D, which does not require
 the calculation or storage 
of Kronecker products, becomes a suitable alternative for
orthogonalization of the MP approach.
\section{Numerical Experiments and Results}
\label{experi}
The viability of using the mixed dictionary described
in Section \ref{dic} to quickly approximate an image by 
either OMP2D or SPMP2D 
follows from its suitability for block processing. This implies to
 divide the image $\vI$ into small blocks, for  independent
approximation.

Without loss of generality blocks are assumed to be square
of size $N_h\times N_h$ pixels. Also for simplicity an image $\vI$  will
be assumed to be the composition of $H$ identical blocks, i.e.,
$$\vI = \cup_{h=1}^{H} \vI_h,$$
where every $\vI_h$ is an intensity array of size ${N_h}\times N_h$,
to be approximated as
\be
I^{K_h}(i,j)= \sum_{n=1}^{K_h}
c^{K_h}(n) \dx_{\ell^x_n}(i)\dy_{\ell^y_n}(j), \quad i,j=1,\ldots N_h.
\ee
In what follows the performance of our dictionary 
based approach is illustrated by   
recourse to numerical experiments.
The sparsity is measured by the Sparsity Ratio (SR)
defined as 
$$\text{SR}=\frac{\text{number of pixels}}{\text{number of coeffients}}= \frac{H N_h^2}{\sum_{h=1}^H {K_h}}.$$

{\bf{General setup}}\\

The experiments have been realized 
in the MATLAB programming environment, on a laptop with
a 2.4GHz Intel Core 2 Duo P8600 processor and 7.7GB
of RAM.

In all the cases the approximation tolerance 
is fixed to produce a sharp PSNR of 45dB for the complete image.  
For such a PSNR  
Mean Structure Similarity index (MSSIM) 
\cite{ssim} with respect to the original image 
is very close to one (MSSIM $> 0.98$ for all the images).

Unless explicitly specified 
the mixed dictionary is the one introduced 
in section Sec.\ref{dic}, i.e. the union of a RDC dictionary, redundancy 
two, and the RDBS dictionary arising by translation of the eight prototypes 
shown in Fig.~2.

The comparison with the DCT and DWT refers to the nonlinear approach 
achieving, by thresholding of the DCT and  DWT coefficients 
respectively, the required PSNR of 45 dB.
The DWT is applied to the whole image using software implementing 
the Cohen-Daubechies-Feauveau 9$/$7 wavelet transform.\\ 

{\bf{Experiment I}}\\

The aim is to evaluate the sparsity of an image representation, 
by the proposed mixed dictionary,
against the size of the blocks partitioning the image. 
The results are  compared with those yielded by the 2D 
version of the DCT and DWT orthogonal transforms.
With this end in mind, numerical experiments were conducted 
using a set of images 
downloaded from the ESO website \cite{esoweb}, 
converted to gray intensity levels, for two resolutions 
(publication and screen size).
We include here full results corresponding to the two 
images in Fig.~3, which 
are good representatives of the range of images in the 
data set tested. 
\begin{figure}
\begin{center}
\includegraphics[width=8cm]{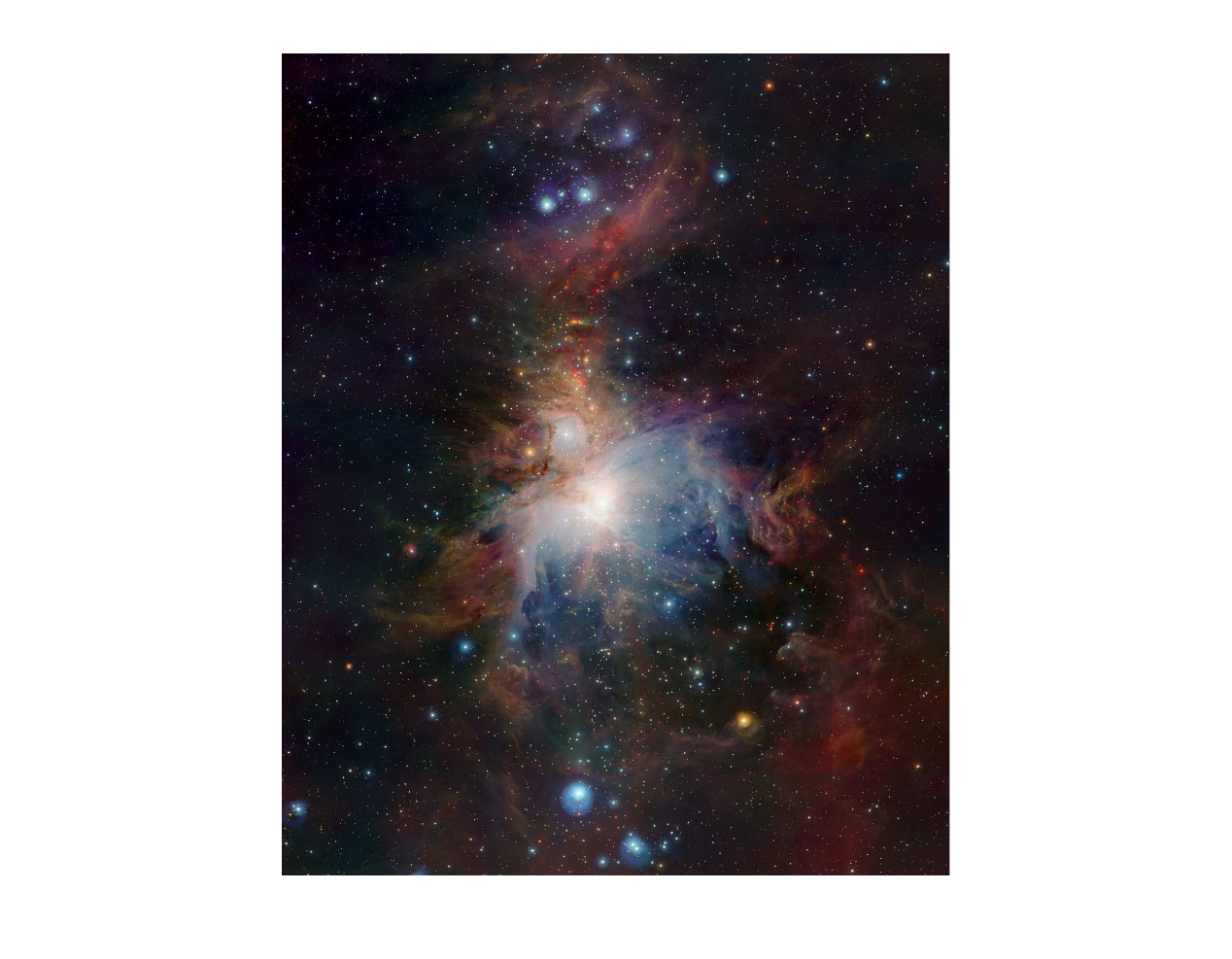}
\hspace{-0.95cm}
\includegraphics[width=8cm]{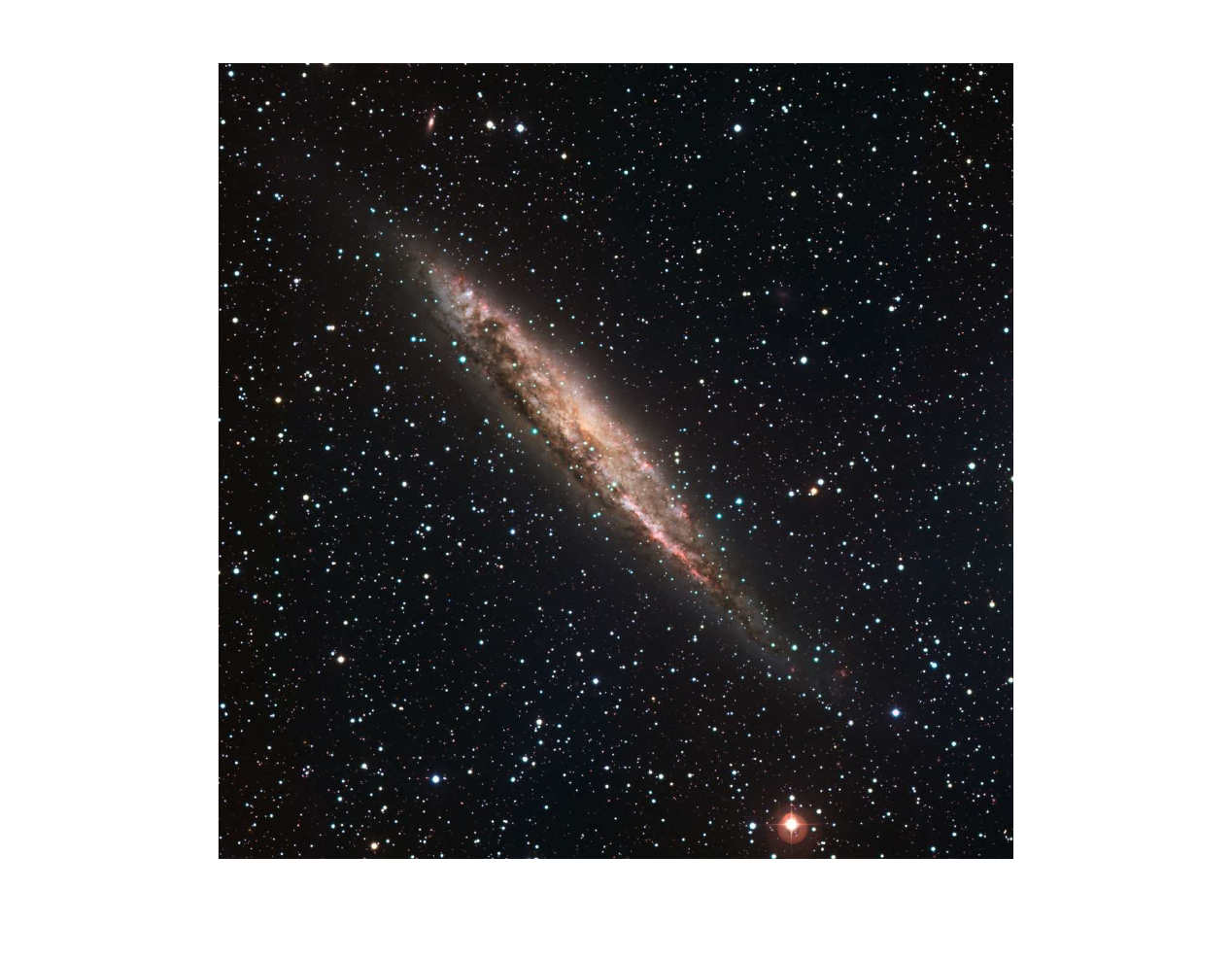} 
\end{center}
\vspace{-0.75cm}
\caption{\small{The first image is the Nebula Orion (Messier 42 
Ref. eso1006) and the second the Spiral Galaxy NGC 4945 
Ref. eso0931). Both images are taken from 
ESO \cite{esoweb} at publication and screen resolutions. The corresponding 
sizes (in pixels) are: 
$4000 \times 3252$ and  $1280 \times 1574$ (nebula)
$4000 \times 4000$ and $1280 \times 1280$ (galaxy).}}
\end{figure}

For a fixed PSNR of 45dB the SR is calculated by partitioning 
the corresponding image into square blocks 
of side length $8,16,24,32,40$ and $48$. 
Fig.~4 depicts the SR obtained, using the mixed dictionaries  
 from  Sec.~\ref{dic}, and  OMP2D, SPMP2D with projection 
 step one (SPMP2D$_1$) and ten (SPMP2D$_{10}$), and the 2D version 
 of MP, for separable dictionaries, that we denote MP2D.
Sparsity is also compared against results for the DCT
(for the same block size) 
and the DWT applied to the whole image.
\begin{figure}
\begin{center}
\includegraphics[width=8cm]{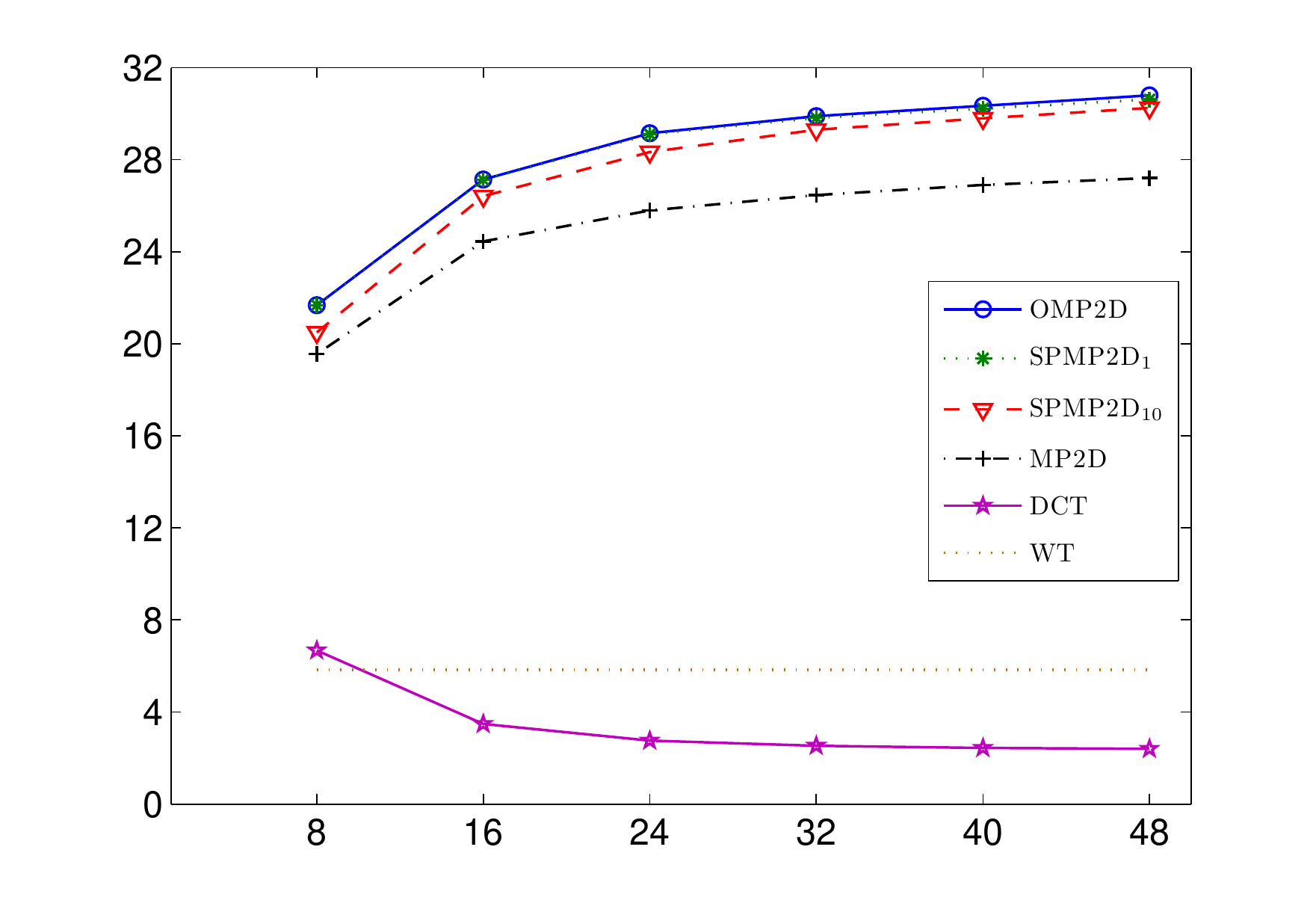}
\includegraphics[width=8cm]{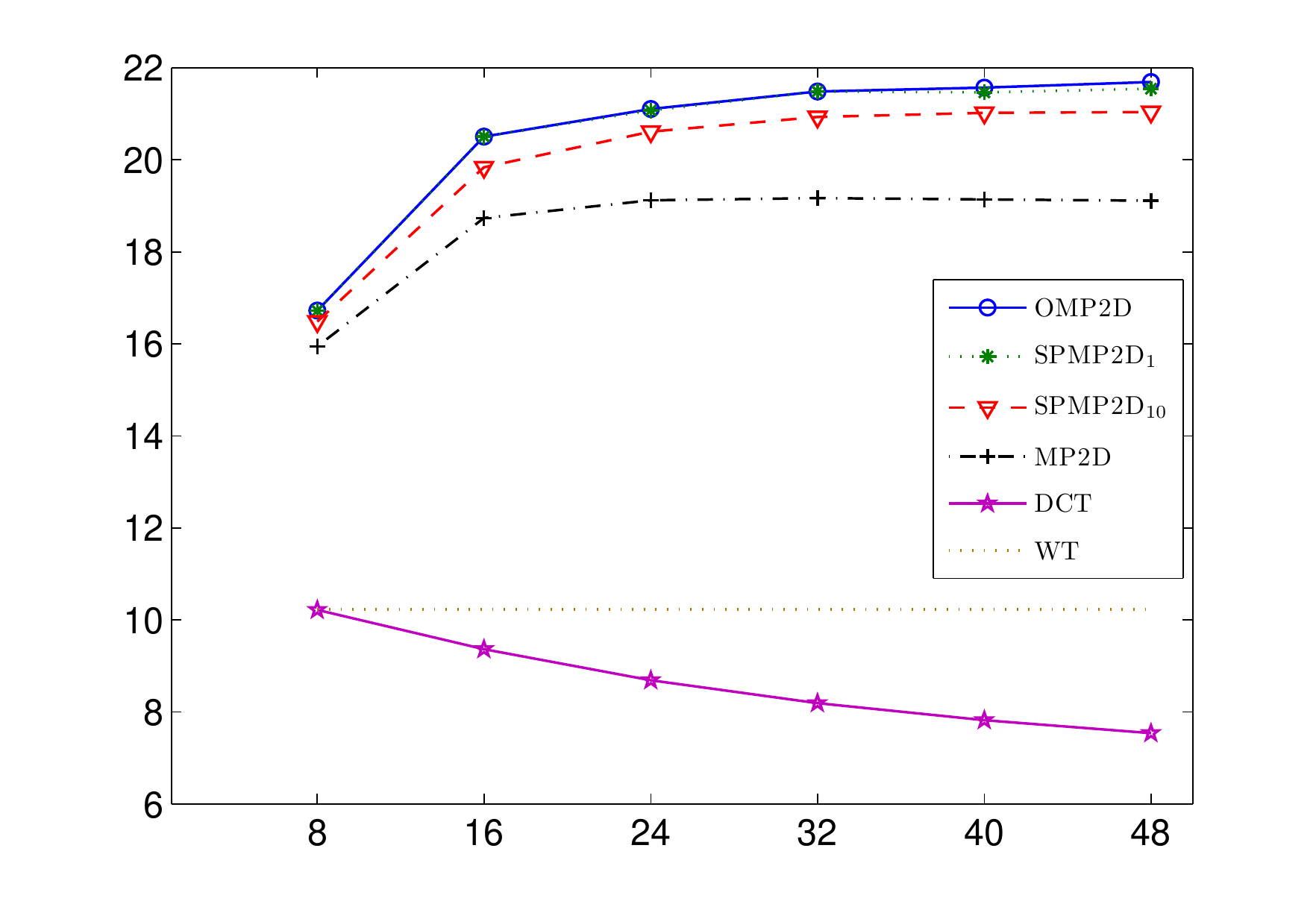}\\
\includegraphics[width=8cm]{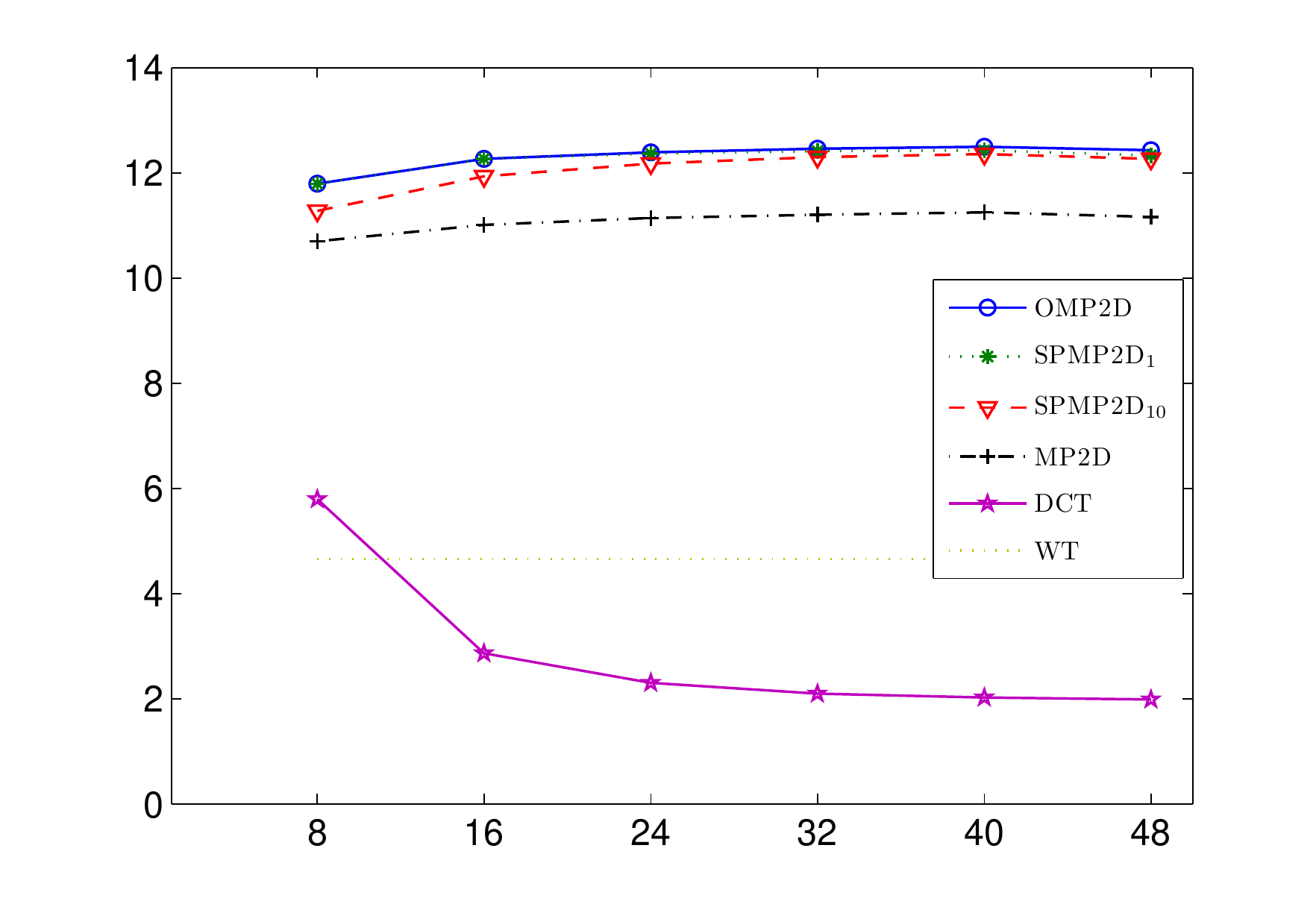}
\includegraphics[width=8cm]{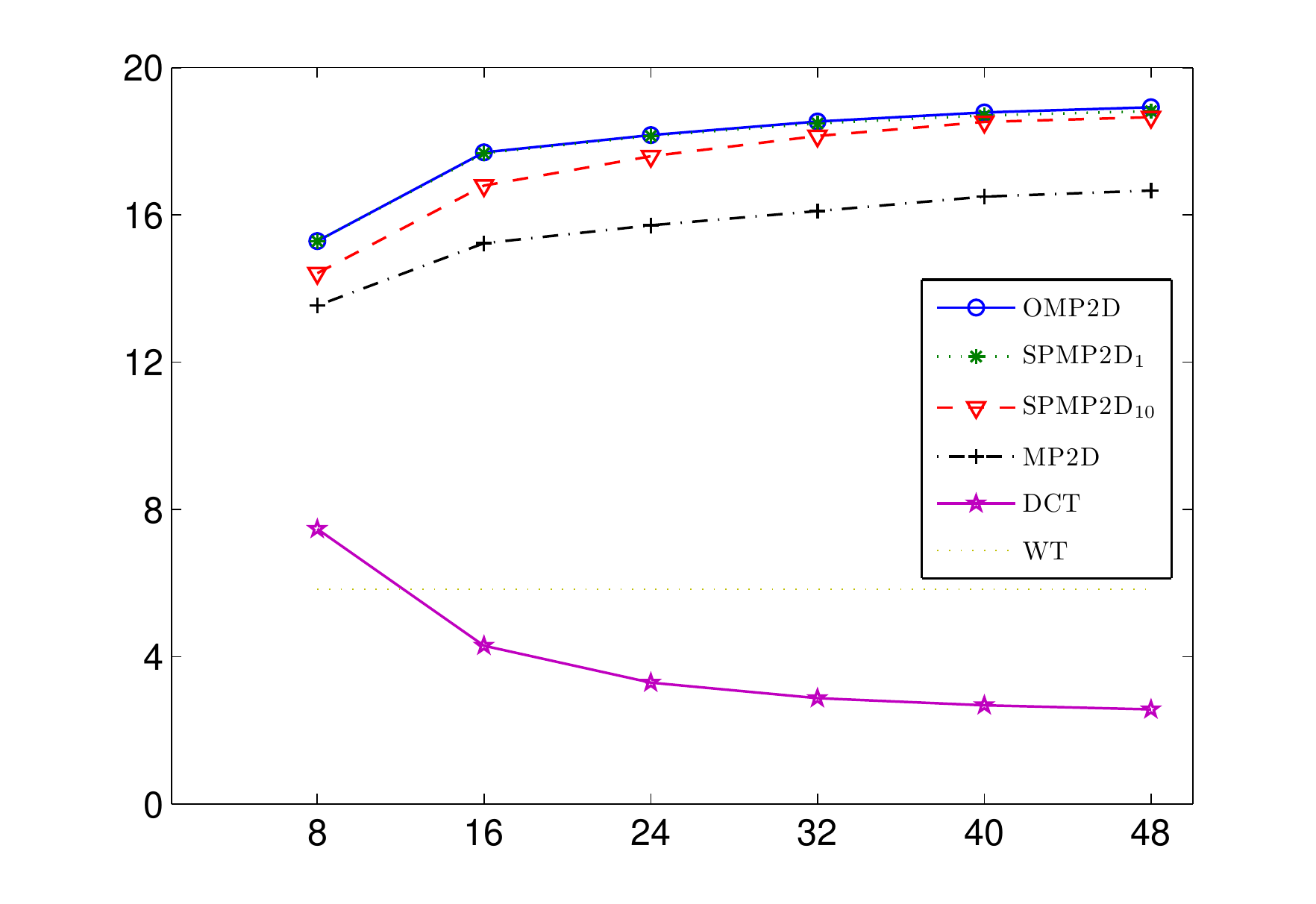}
\end{center}
\vspace{-0.75cm}
\caption{\small{SR vs partition of side length $8,16,24,32,40$, and 48
yielded by 
OMP2D, SPMP2D with projection
step one (SPMP2D$_1$) and ten (SPMP2D$_{10}$), MP2D, 
and the DCT. The constant dotted line corresponds to the DWT result  
and is plotted only for visual comparison, since the DWT is applied 
to the whole image.
The left and right graphs correspond to the nebula 
and galaxy of Fig.~{{3}}, respectively. The top graphs 
correspond to the higher resolution 
$4000 \times 3252$ pixels and $4000 \times 4000$ pixels respectively. 
The bottom 
graphs correspond to the lower resolution $1280 \times 1574$ and
 $1280 \times 1280$ pixel images respectively.}}
\label{esoresults}
\end{figure}
The points joined with different lines in the top left graph 
of Fig.~{4}
show  results for the  first image of Fig.~{3} at the
higher resolution
($ 4000 \times 3552$  pixels). The results corresponding 
to the second image of Fig.~3, also 
at the higher resolution 
($4000 \times 4000$ pixels), are shown in the top right 
graph of Fig.~4. 
The bottom left and right  graphs  depict the
same information as the top graphs but correspond to the 
screen size resolution of the images 
(size $ 1280 \times 1574$  pixels and  size $1280 \times 1280$ pixels
respectively).\\

{\bf{Discussion of results}}\\

Let us start by highlighting the fact that the results of  
Fig.~4 illustrate a \underline{massive} 
gain in sparsity yielded by the dictionary approach,  
in comparison to the DCT and DWT.

A clear feature in the results corresponding to the higher 
resolution images (top graphs of Fig.~4) 
is that, as opposed to the results for the DCT
(decreasing curve in all the graphs of Fig.~4),
the SR yielded by the dictionary approach increases 
with the block size, rapidly up to some value. 
For most images in the data set, 
and in particular for the two images  of Fig.~3, 
block size $16 \times 16$ yields the best
 trade off between sparsity and processing time (c.f. Tables 1).

The two  bottom graphs confirm that, 
as should be expected, for the fixed 
PSNR of 45dB sparsity decreases with respect to the previous 
resolution and the block size 
has less relevance.

Notice that for the lower resolution image of the 
nebula, the SR shown in the bottom left graph of Fig.~4 
becomes almost uniform for block sizes larger 
than $16 \times 16$ pixels. 
For the lower resolution image of galaxy the SR shown 
in the bottom right graph of Fig.~4 increases 
with the block size, but much less than for the same image at higher 
resolution (top right graph).
\begin{table}
\label{tab1}
\begin{center}
\begin{tabular}{| r || r |r|| r|r || r|r || r|r|| }
\hline
Block size & \multicolumn{2}{|c||}{OMP2D} & \multicolumn{2}{|c||} {SPMP2D$_1$} & 
\multicolumn{2}{|c||} {SPMP2D$_{10}$}&
\multicolumn{2}{|c||}{ MP2D}\\ \hline \hline
               &SR &secs &SR&secs&SR&secs &SR &secs \\ \hline
$8   \times 8 $ & 21.69 &  51   &  21.69 &    61   & 20.52  &   60   & 19.55&   56 \\ \hline
$16  \times 16$ & 27.63 &  98   &  27.61 &   115   & 26.41  &   99   & 24.49&   93 \\ \hline
$24  \times 24$ & 29.15 &  233  &  29.08 &   209   & 28.28  &  200   & 25.79&  205 \\ \hline
$32  \times 32$ & 29.97 &  506  &  29.88&    392   & 29.30  &  382   & 26.46&  387 \\ \hline
$40  \times 40$ & 30.36 & 1065  &  30.24 &   666   & 29.80  &  640   & 26.90&  660 \\ \hline
$48  \times 48$ & 30.78 & 2041  &  30.60&   1055   & 30.25 &  1015   & 27.20 & 1032 \\ \hline  \hline
$8 \times   8 $ & 16.85 &  79   & 16.85  &    85  & 16.49  &    89   & 15.95&   79 \\ \hline
$16\times  16 $ & 20.51 &  163  & 20.50 &    185  & 19.84  &   154   & 18.73&  147 \\ \hline
$24 \times 24 $ & 21.27 &  413  & 21.23  &   362  & 20.61  &   354   & 19.12&  328 \\ \hline
$32 \times 32 $ & 21.59 &  916  & 21.52 &    694  & 20.93  &   666   & 19.16&  653 \\ \hline
$40 \times 40 $ & 21.70 &  1989  & 21.59 &  1494  & 21.02  &  1145   & 19.13&  1139 \\ \hline
$48 \times 48 $ & 21.68 & 4031  & 21.53 &   2477  & 21.04  &  1919   & 19.11&  1853 \\ \hline 
\end{tabular}
\caption{\small{SR and execution time, in secs, for approximating  a 
 complete image with different approaches
 and different sized 
 blocks partitioning the image.
The top half of the table corresponds to the results for the
nebula at publication size resolution ($4000\times 3252$) 
pixels. The  bottom half contains  the
results for the galaxy image at the equivalent resolution 
($4000 \times 4000$ pixels).}}
\end{center}
\end{table}
Table 1 compares the time (average of 5 independent runs)
spent by the methods considered here, using the 
proposed dictionary, vs the size of the  blocks partitioning the 
image. As can be observed,   
OMP2D implemented as described in Sec.~{\ref{omp2d}} is slightly 
faster than 
SPMP2D with projection step one, up to block size $24 \times 24$, 
and becomes noticeable slower beyond that block size. 
This behavior is 
not a consequence of mathematical complexity, which as discussed in 
Sec {\ref{omp2d}} in the best scenario are at most of the same order, 
but as a consequence of storage
requirements. 
As the block size increases, 
the poor execution time scaling for OMP2D 
is a result of the additional memory required, over SPMP2D, 
for the storage  of matrices $\vB$ and $\vW$
(c.f. \eqref{BW}).
Another interesting feature is that the results of SPMP2D 
with projection step larger than one do not improve 
the processing time significantly.\\

{\bf{Experiment II}}\\

This experiment comprises a data set composed of 
fifty five images at screen size resolution,
all of them in the category of nebulae, galaxies, and stars,
taken from the top 100 images on the HST website \cite{hubweb}.
Table~2 displays the average SR for the set, denoted as $\mur$
as well as the average processing time, $\overline{t}$, per 
image in the set,  using OMP2D, SPMP2D$_1$, SPMP2D$_{10}$, and 
MP2D, with partitions of square blocks 
of sides $8,16,24,32,$ and 40. 
The average size of the images in the 
set is $1264\times 1194$ pixels.
\begin{table}
\label{tab3}
\begin{center}
\begin{tabular}{| l || c |c|| c|c || c|c || c|c |}
\hline
Block size & \multicolumn{2}{|c||}{OMP2D} & \multicolumn{2}{|c||} { SPMP2D$_1$} &  \multicolumn{2}{|c||}  { SPMP2D$_{10}$}&
\multicolumn{2}{|c||}{MP2D} \\ \hline \hline
               &$\mur$ &$\tr$ &$\mur$&$\tr$&$\mur$&$\tr$ &$\mur$ &$\tr$  \\ \hline
	       $8 \times 8 $  & 12.36 & 11.26 &  12.46 & 13.70 &   12.18 &  12.2 &  11.72 &12.37\\ \hline
	       $16 \times 16$ & 14.35 & 38.11 &  14.42 & 45.23 &   14.13 &  34.22 & 13.21 &28.52 \\ \hline
	       $24 \times 24$ & 14.94 & 113.27&  14.96 & 111.44 &  14.74 & 91.11& 13.59 &70.66\\ \hline
	       $32 \times 32$ & 15.23 & 326.09&  15.22 & 237.79 &  15.05 & 207.47&13.78 &134.73\\ \hline
	       $40 \times 40$ & 15.36 & 839.47&  15.31 & 447.20 &  15.17 & 397.81&13.83 &239.10\\ \hline
	       \end{tabular}
\caption{\small{Average SR ($\mur$) and average processing time
   ($\tr$) per image (in secs)
 for approximating,  up to a PSNR of 45 dB, a  set of 55 images
 from the HST website. Both quantities are displayed  against
 the block size partitioning the images. 
 The average size of the images in the
 set is $1264\times 1194$ pixels.
 Note: the average times per image are also the average 
 of 5 independent runs for each given block size.}}
\end{center}
\end{table}

Now we are also interested in testing the
proposed mixed dictionary against other possible mixed dictionaries.
Preliminary experiments have shown that
all combination of dictionaries containing a RDC (with redundancy two) 
component perform better than combinations 
without this component. Considering the preliminary tests we compare 
here the results obtained with the 
dictionaries of Sec.~\ref{dic}, against other mixed dictionaries 
containing a RDC component. The Euclidean basis is 
also included in all dictionaries. The 
 comparison is carried out with respect to 
 the remaining localized atoms.
The RDBS dictionary is replaced by another one
constructed in an equivalent manner using the prototypes 
in the top graphs of Fig.~5. The atoms of 
support 2,4,6, and 8, 
represented in the top right graph of 
Fig.~5, are discretized versions of Haar wavelets.
The other prototypes are discretized versions of 
the continuous wavelets given in \cite{Dau92}, the form of which 
is very similar to the Mexican Hat wavelet. 
Three fractional scaling parameters were used  
to produce discrete  wavelets of 
support $3, 5$, and $7$, represented in the top left graph of
Fig.~5.
We call this dictionary Redundant Discrete Wavelet (RDW) dictionary. 
For further comparison we constructed a random dictionary from
normal distributed random atoms of support equal to the atoms 
of the dictionary we are testing against.
We call such a dictionary Redundant Random (RR) dictionary. The 
prototypes corresponding to a particular realization of the 
random shapes are shown in the bottom graphs of Fig.~5.
\begin{figure}
\begin{center}
\includegraphics[width=7cm]{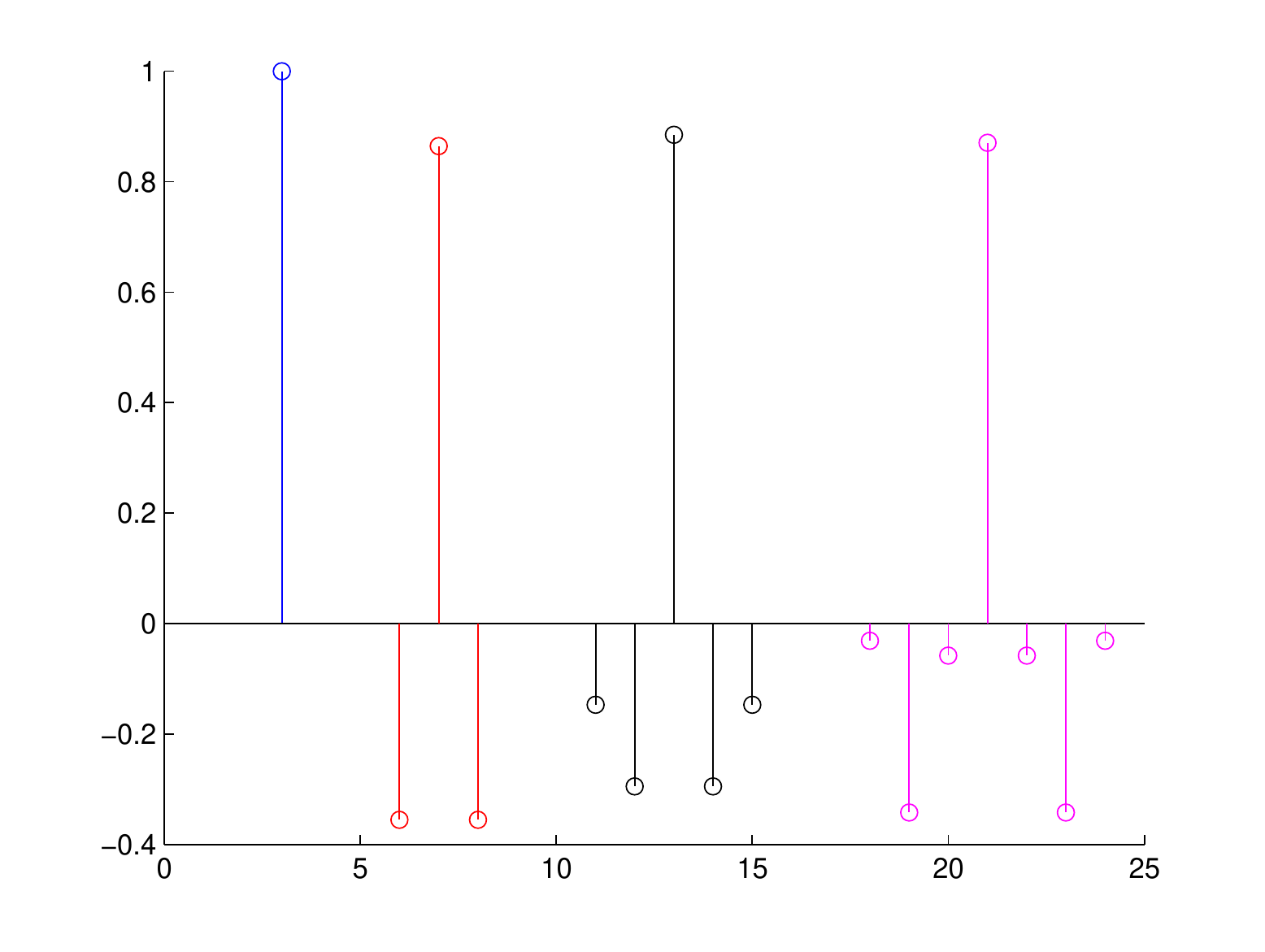}
\includegraphics[width=7cm]{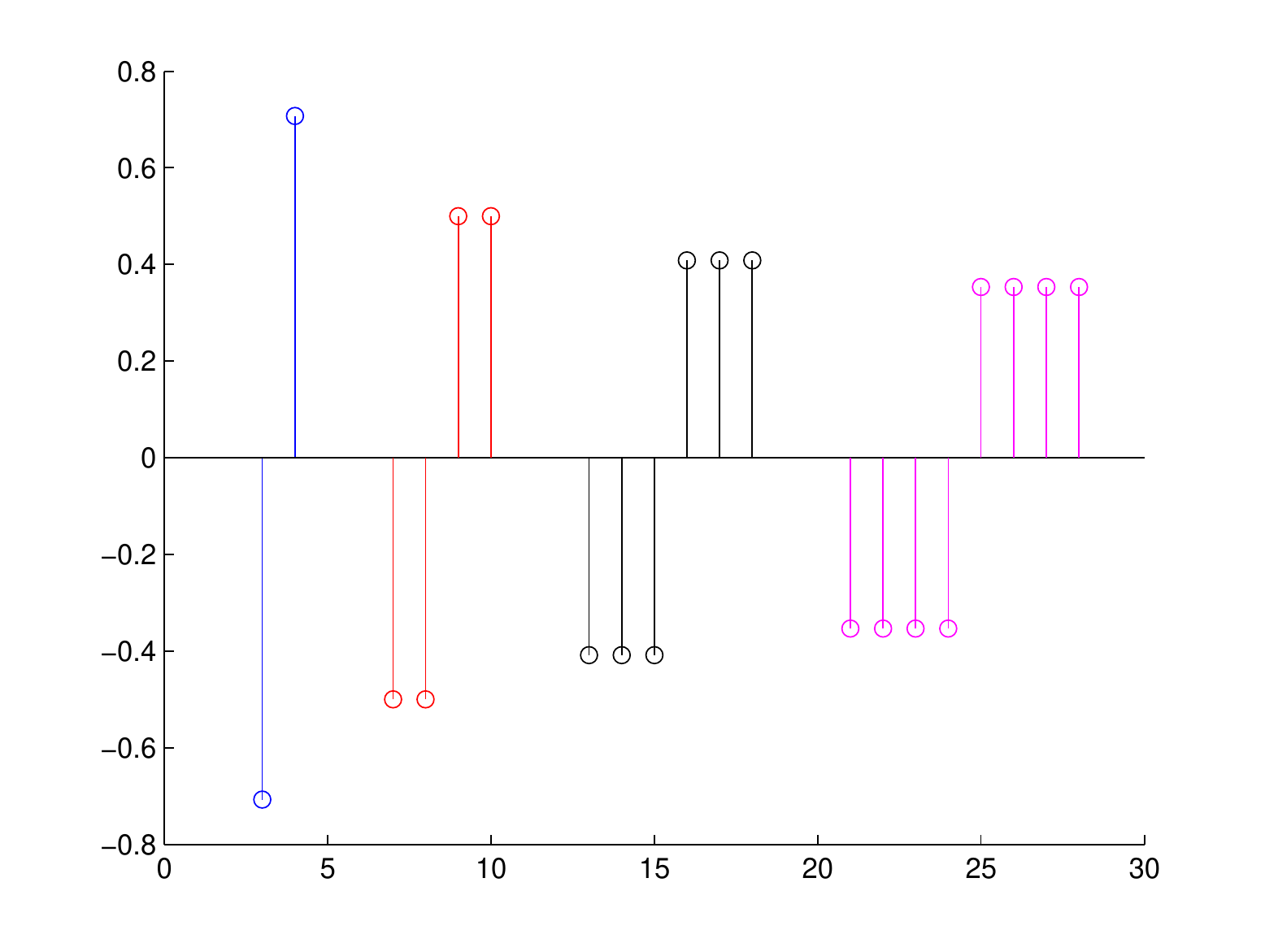}\\
\includegraphics[width=7cm]{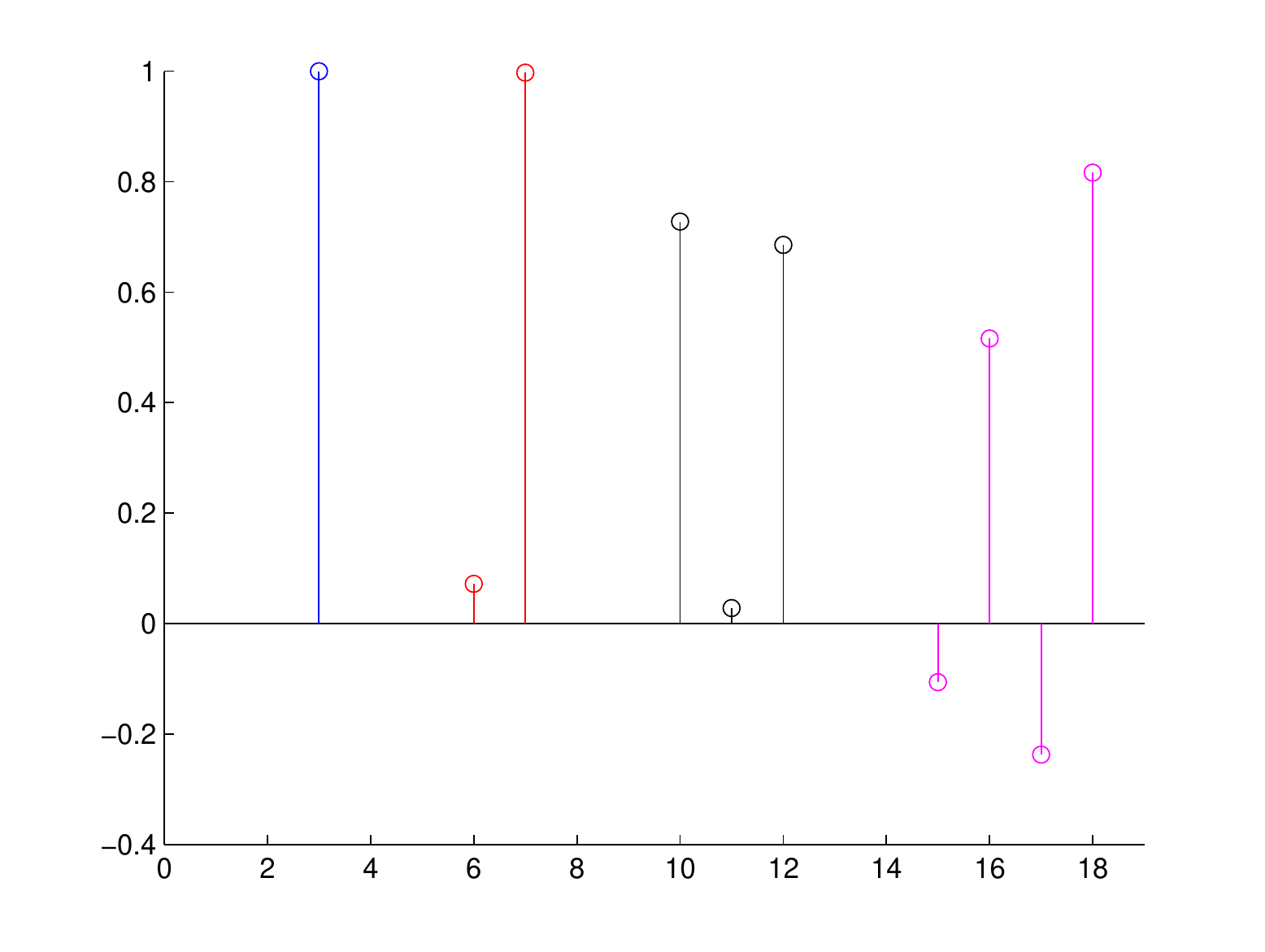}
\includegraphics[width=7cm]{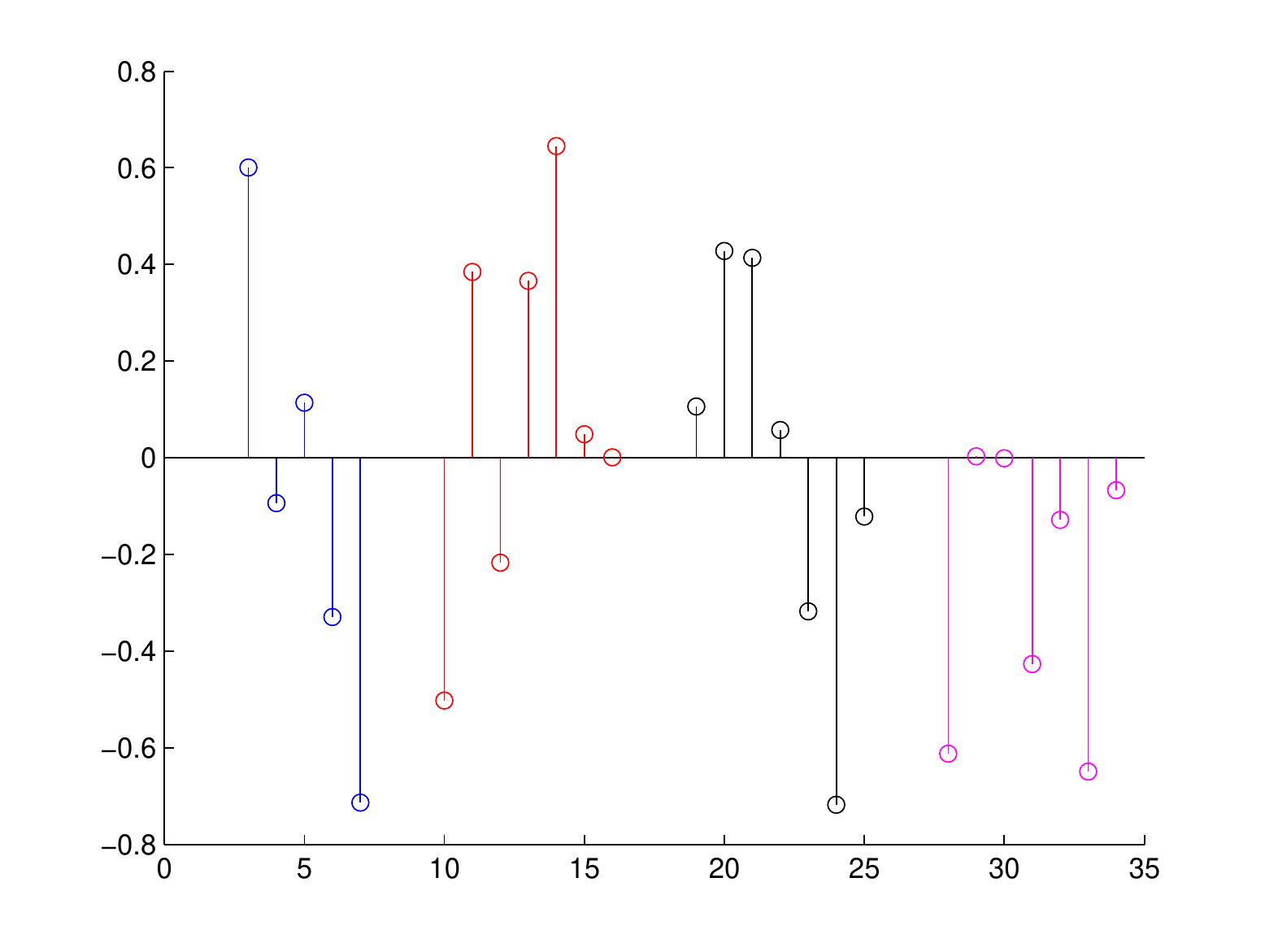}
\end{center}
\vspace{-0.75cm}
\caption{\small{The two top graphs are 
the prototype atoms defining the RDW dictionary. 
The two bottom graphs  are prototype atoms of random shape
defining a realization of the RR dictionary.}}
\end{figure}
For the experiment, 
five different realizations of a RR dictionary were 
tested. 
All the realizations produce similar results. The results 
displayed in Table~\ref{WHRR} correspond to the average of 
the five realizations. 
Thus, the corresponding sparsity ratio $\omur$ 
is a double average. Namely, the average sparsity ratio $\mur$ 
for the set of fifty five images and the average of this quantity $\omur$ 
 with respect to the five realizations of 
the RR dictionary. As already mentioned the $\mur$ does not 
depend  significantly on the dictionary realization (the standard deviation 
of $\omur$ with respect to the five random realizations  is, 
for all the block sizes, 
less than $4\%$ of the given $\omur$  values). 
The standard deviation with respect to the fifty five images is 
also the average $\osigr$ corresponding to the five 
realization of the RR dictionary. 
Table ~\ref{WHRR} also shows the $\mur$ and $\sigr$
for mixed dictionaries corresponding 
to the components RDBS and RDW and the corresponding 
standard deviations.  
All the dictionary results in Table \ref{WHRR}
are obtained using the OMP2D method for
block sides $8, 16, 24$ and $32$.
The results from the DCT and DWT are also displayed.\\

{\bf{Discussion of Results}}\\

This experiment confirms  statistically the gain in 
sparsity obtained with the proposed dictionaries with 
respect to  the DCT and DWT. It also confirms that the  
proposed approach SPMP2D$_1$ is a faster option for the implementation of 
OMP2D  when the image partition is larger than 
$24\time 24$.

It is clear from Table \ref{WHRR}  that, while the RDW and RR 
dictionary components produce comparable results, 
the differences with the RDBS component 
are significant, specially for the larger  
partition sizes. However, 
comparison with the $\mur$  yielded by the DCT and DWT 
leads to the conclusion  that 
it is the combination of a RDC dictionary with 
well localized atoms of arbitrary shape which 
yields a significant improvement in the sparsity of high quality
approximations of astronomical images. Atoms 
 of particular shape, such as 
the prototypes in the RDBS component improve sparsity even further. 
\begin{table}
\label{tab2}
\begin{center}
\begin{tabular}{| l || c |c|| c|c || c|c || c|c || c|c||}
\hline
Block size & \multicolumn{2}{|c||}{RDCT-RDBS} & \multicolumn{2}{|c||} {RDCT-RDW} &  \multicolumn{2}{|c||}  {RDCT- RR}& 
\multicolumn{2}{|c||}{DCT}& \multicolumn{2}{|c||} {DWT} \\ \hline \hline
               &$\mur$ &$\sigr$ &$\mur$&$\sigr$&$\omur$&$\osigr$ &$\mur$ &$\sigr$ &$\mur$ &$\sigr$ \\ \hline
	$8 \times 8 $ & 12.36  & 6.2   &  11.28   & 5.7 & 10.99 & 5.5 & 7.94 & 4.5& 6.39 & 4.9\\ \hline
	$16 \times 16$ & 14.35  & 8.9   &  12.38   & 7.5 & 12.07 & 7.2 & 6.13 & 4.4& 6.39 & 4.9\\ \hline
	$24 \times 24$ & 14.94  & 9.7   &  12.52   & 7.8 & 12.29 & 7.5 & 5.66 & 4.3& 6.39 & 4.9\\ \hline
	$32 \times 32$ & 15.23  & 10.0  &  12.56   & 7.9 & 12.42 & 7.6 & 5.39 & 4.2& 6.39 & 4.9\\ \hline  
\end{tabular}
\caption{\small{Mean ($\mur$) and variance ($\sigr$) 
of the SR 
obtained with different mixed  
dictionaries,  by partitioning the 
images into blocks of size 8, 16, 24  and 32 and applying the 
OMP2D approach with RDCT-RDBS, RDCT-RDW and RDCT-RR dictionaries. 
The results from the  approximation with the transforms DCT and WT are 
also shown.}}
\label{WHRR}
\end{center}
\end{table}
\section{Conclusions}
\label{conclu}
Sparse representation of astronomical images has been 
considered. A mixed dictionary composed of a RDC 
component and a RDBS component was proposed. Using a data set 
of fifty five astronomical images in the  category of nebulae, galaxies, 
and stars, the dictionary was shown to be suitable for  
sparse representation of that class of images. From the 
experiments involving atoms of different shapes one can 
assert that the combination of a RDC component with a component of
localized atoms of different support yields the 
most important gain in sparsity, with respect to results from 
the popular transforms DCT and DWT. Nevertheless, 
the proposed particular shape of the RDBS component  
 represents an advantage 
over other possible atoms of equal support, and yields an
impressive sparsity gain over DCT and DWT results.

The fact that the proposed dictionary is suitable 
for block processing by the selection 
technique OMP2D makes the resulting approach very effective in 
terms of processing time. For the data set 
 characterized by an average sparsity ratio of $12.6$ 
 (with standard deviation $6.2$) 
the processing time, for partition size $8 \times 8$, 
is only $11.26$ secs per image of average size of $1264\times 1194$
pixels. 
This should be appreciated taking into account that the 
time refers to executing a C++ MEX file in 
a MATLAB environment, using a small laptop with the 
specification details given in Sec.~\ref{experi}.

Specially for high resolution 
images, sparsity may significantly increase with the 
partition size. In order to handle 
these cases, a greedy 
strategy taking full advantage of the separability 
property of the proposed dictionaries was considered. 
 The approach 
 has been termed SPMP2D, because it allows to orthogonalize the 
 seminal MP technique by self-projections. Through the experiments 
 the technique was established as a  convenient 
 alternative to OMP2D, when the 
 latter scale badly due to storage demands, or the storage capacity
 is not available. 
 
Finally we would like to highlight that, even though 
for small partitions the approach is fast for sequential computing, 
the possibility of its parallel  implementation is only a question of 
 resource availability. The approach obeys
a scaling law by independently processing the 
blocks partitioning the image. Hence, a straightforward  
 multiprocessor implementation 
would reduce the processing time of the sequential implementation 
by a factor approximatelly equal to the number of processors.  

The results presented in this Communication 
are truly encouraging and we feel confident that the approach will benefit 
applications relying on sparse representation of digital images.  
\subsection*{Acknowledgements}
Support from EPSRC, UK, grant (EP$/$D062632$/$1) is acknowledged.
The MATLAB files and C++ MEX files for implantation of 
the OMP2D and SPMP2D methods are available at \cite{webpage2}.
We are grateful to ESO and HST for making  publicly available the 
images that have been used for the experiments.
\bibliographystyle{elsart-num}
\bibliography{revbib}
\end{document}